%% file: main.tex
\documentclass[twocolumn,prb,amsmath,amssymb,superscriptaddress]{revtex4}
\usepackage{xcolor}
\usepackage[utf8]{inputenc}
\usepackage{amsmath}
\usepackage{mathtools}
\usepackage{amsfonts}
\usepackage{amssymb}
\usepackage{graphicx}
\usepackage{wrapfig}

\usepackage{MnSymbol}
\usepackage{braket}
\usepackage{comment}
\usepackage{array}
\usepackage{float}
\usepackage{bm}
\usepackage{bbm}
\usepackage{dsfont}
\usepackage{cellspace}
\usepackage{standalone}
\usepackage{tikz}
\usepackage[colorlinks]{hyperref}

\allowdisplaybreaks

\begin{document}
\title{
Spin and pair density waves in 2D altermagnetic metals}

\author{Nikolaos Parthenios}
\affiliation{Max Planck Institute for Solid State Research, D-70569 Stuttgart, Germany}
\affiliation{School of Natural Sciences, Technische Universit\"at München, 85748 Garching, Germany}

\author{Pietro M. Bonetti}
\affiliation{Department of Physics, Harvard University, Cambridge MA 02138, USA}
\affiliation{Max Planck Institute for Solid State Research, D-70569 Stuttgart, Germany}

\author{Rafael Gonz\'{a}lez-Hern\'{a}ndez}
\affiliation{Departamento de F\'{i}sica y Geociencias, Universidad del Norte, Km. 5 V\'{i}a Antigua Puerto Colombia, Barranquilla 081007, Colombia}

\author{Warlley H. Campos}
\affiliation{Max Planck Institute for the Physics of Complex Systems, N\"othnitzer Str. 38, 01187 Dresden, Germany}

\author{Libor Šmejkal}
\affiliation{Max Planck Institute for the Physics of Complex Systems, N\"othnitzer Str. 38, 01187 Dresden, Germany}
\affiliation{Max Planck Institute for Chemical Physics of Solids, N\"othnitzer Str. 40, 01187 Dresden, Germany}
\affiliation{Institute of Physics, Czech Academy of Sciences, Cukrovarnick\'a 10, 162 00, Praha 6, Czech Republic}

\author{Laura Classen}
\affiliation{Max Planck Institute for Solid State Research, D-70569 Stuttgart, Germany}
\affiliation{School of Natural Sciences, Technische Universit\"at München, 85748 Garching, Germany}

\begin{abstract}
Altermagnetism, a recently proposed and experimentally confirmed class of magnetic order, features collinear compensated magnetism with unconventional 
d-, 
g-, or 
i-wave spin order. Here, we show that in a metallic 2D 
d-wave altermagnet with combined two-fold spin and four-fold lattice rotational symmetry 
$[C_2||C_4]$, secondary instabilities can arise. 
Using an unbiased functional renormalization group approach, we analyze the weak-coupling instabilities of a 2D Hubbard model with a preexisting altermagnetic order inspired by our ab initio electronic structure calculations of realistic material candidates from V$_2$X$_2$O (X = Te, Se) family. We identify two distinct spin density wave (SDW) states that break the underlying altermagnetic 
$[C_2||C_4]$
 symmetry. Additionally, we find spin-fluctuation-induced instabilities leading to a singlet 
d-wave superconducting state and an unconventional commensurate pair density wave (PDW) state with extended 
s-wave and spin-triplet symmetry.
We establish a general criterion for the unusual exchange %
statistics for these pair density waves and characterize their excitation spectrum, which exhibits Bogoliubov Fermi surfaces or nodal points depending on the gap size.
\end{abstract}

\maketitle

\section{Introduction}
An unconventional form of collinear magnetic order, dubbed altermagnetism, was recently theoretically proposed to be ubiquitous in many materials \cite{Smejkal2022Sep,Smejkal2022Dec,Bai2024Sep} and experimentally  confirmed in MnTe and CrSb \cite{Krempasky2024Aug,Lee2024Jan,Osumi2024Mar,Reimers2024Mar,Jiang2024Aug,Amin2024Dec,Li2024May}.  This particular state differs from  ferro- and antiferromagnets, as it breaks time reversal symmetry (TRS) in the form of d, g, or i-wave magnetization in direct space with zero net magnetization \cite{Smejkal2022Sep,Smejkal2020Jun}. In momentum space, the altermagnetic order also exhibits the corresponding even-parity-wave spin splitting. 
The experimentally confirmed  altermagnets exhibit  two spin-sublattices, which are not mapped onto each other by translation or inversion as in conventional antiferromagnets, but instead by a real-space rotation. As a consequence, altermagnets possess distinct electronic, magnetic, and topological properties, which manifests in unconventional spin, charge and optical responses \cite{Fang2023Oct,Gonzalez-Hernandez2021Mar,Hariki2024Apr,Parshukov2024Mar,GonzalezBetancourt2023Jan,Smejkal2022NatMat,Takahashi2025Feb,Fernandes2024Jan,Rao2024Jul,Attias2024Sep,Roig2024Dec,Sourounis2024Nov,Banerjee2024Jul,Bai2024Sep,Smejkal2020Jun,Smejkal2022Dec}.

The most common mechanism that captures altermagnetism is due to anisotropies in the crystal lattice \cite{Smejkal2022Sep,Smejkal2022Dec,Fernandes2024Jan,Guo2023Mar} and as such can be readily captured by ab initio methods such as density functional theory (DFT). In this scenario, the local anisotropy of the crystal induces local anisotropic spin densities in the magnetic phase. 

The altermagnetic (AM) transition can happen at high temperatures and reach spin splittings on the order of eV similar to ferromagnets because of its exchange origin. This offers the exciting opportunity to realise quantum many-body states as secondary instabilities of AM metals. They can be expected to be of unconventional type because they arise out of an unconventional  parent state.

While many theoretical studies focused on the mechanism behind the origins of an AM instability \cite{Durrnagel2024Dec,PhysRevLett.132.263402,PhysRevLett.132.236701,yu2024coincidentVH,PhysRevB.110.144412,Liu2022Apr,Leeb2024Jun,Maier2023Sep,Roig2024Feb}, analyses of secondary instabilities are limited. Possible superconducting states in AM metals were discussed, including finite-momentum and topological superconductors \cite{sim2024PDW,PhysRevB.108.184505,Zhang2024finite,bose2024tJ,PhysRevB.110.L060508,PhysRevB.110.024503,Brekke2023Dec,Krempasky2024Aug} due to the lifted Kramers spin degeneracy . Yet, an understanding of their pairing mechanisms possibly from repulsive interactions remains an open problem, despite the importance of the relation between magnetism and superconductivity. Interestingly, AM was also recently theorized to appear as possible interaction-induced instability in cuprates \cite{Li2024Dec}. 

A paradigmatic model for the study of competing magnetic and superconducting orders is the 
square-lattice Hubbard model. 
It underlines the importance of single-particle features such as Van Hove (VH) singularities and Fermi-surface nesting for Fermi-surface instabilities and a weak-coupling pairing scenario. Realization of an analogous model in an AM promises a similarly rich phase diagram with additional, unexplored effects due to the AM splitting. For example, a secondary spin density wave was indicated in the AM KV$_2$Se$_2$O\cite{Jiang2024Aug}.

In the present manuscript, we 
explore the effects of electronic correlations in a square lattice Hubbard model where an AM phase has already been formed, effectively acting as a parent state. We 
employ DFT and functional renormalisation group (FRG) calculations, which allows us to connect the search for suitable candidate materials and the unbiased determination of many-body ground states. 
We present a comprehensive phase diagram in terms of model parameters and make connections with material realizations. We furthermore draw parallels with the well-studied Hubbard model on the square lattice without altermagnetism to understand the qualitative impact on the nature and mechanisms of interaction-driven instabilities. We find instabilities towards distinct density waves, due to the spin splitting and nesting of the Fermi surfaces. In addition, we identify a fluctuation-induced instability to a singlet $d$-wave superconducting state, as well as a novel finite-momentum triplet pairing state enabled by the unconventional splitting of the single-particle bands due to the presence of altermagnetism.    \\

\section{2D Altermagnetic metal material candidates}
\label{sec:Materials}

\begin{figure}[t!]
    \centering
    \includegraphics[width=0.95\linewidth]{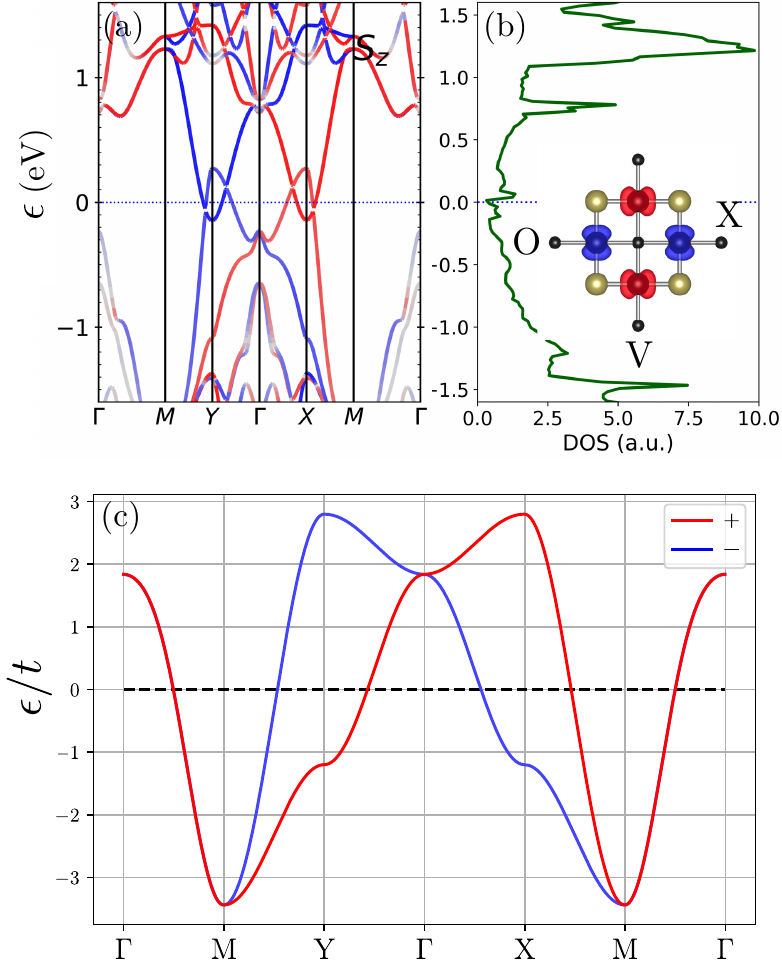}
     \caption{\textbf{Electronic structure of 2D AM metal.} (a) The predicted band structure of $\mathrm{V}_2\mathrm{Te}_2\mathrm{O}$ with $f_{x^2-y^2}$ altermagnetic splitting and (b) its density of states. The inset corresponds to the atom arrangement in the unit cell. (c) For the effective model in our calculations we used $t=-\frac{2}{3},\,t^\prime=-\frac{1}{5}, \lambda=-1$, which provides an adequate description for the band structure for $E-E_F\in[-0.25,0.6]$. Color represents the spin quantum number of the bands.}
    \label{fig:DFT}
\end{figure}

We are interested in candidate materials with favorable properties for a secondary transition inside the AM phase. Typically, Stoner-like instabilities in any ordering channel can arise in a Fermi liquid for strong enough interactions when the dimensionless interaction $u\rho_0\gtrsim 1$. Here, $u$ is the coupling in the ordering channel and $\rho_0$ the density of states (DOS) at the Fermi level. We see that transition temperatures can be boosted when the DOS exhibits a peak. Thus, quasi-2D systems are particularly interesting for us because it is guaranteed that they possess a Van Hove singularity with a logarithmic divergence of the DOS. We note in passing that high-order Van Hove singularities can potentially increase transition temperatures even more \cite{HOVHReview2024}.

Canonical or even approximate Fermi surface nesting, where the dispersion $\xi_\mathbf{k+q}\approx-\xi_\mathbf{k}$ for a finite $\mathbf{q}\neq0$, is another feature beneficial for interaction-induced instabilities, leading to spin or charge density waves.  
The presence of their fluctuations is also 
advantageous for electronic pairing mechanisms, such as a Kohn-Luttinger or spin-fluctuation mechanism. If the interaction becomes attractive, a pairing instability arises whose type depends on the mediating fluctuations and the Fermi surface. Due to the unconventional spin splitting in an altermagnet, they bear the potential for triplet and finite-momentum pairing. The latter can be boosted by nesting in the Cooper channel $\xi_{\mathbf{-k+q}}=\xi_{\mathbf{k}}$, $\mathbf{q}\neq0$.

Based on these general considerations, we can expect strongly competing orders in a Van Hove scenario with nesting. 
Thus, we are looking for quasi-2D AM metals which exhibit Van Hove singularities and/or approximate nesting near the Fermi level. 

As a feasible candidate, we have identified a monolayer of V$_2$Te$_2$O. V$_2$Te$_2$O exhibits the electronic structure shown in Fig. (\ref{fig:DFT}a-b) favourable for the generation of secondary instabilities. We show the calculated magnetisation density in the unit cell of V$_2$Te$_2$O in the inset of Fig. \ref{fig:DFT}b. The opposite local magnetizations of the V atoms are related by a combined $[C_{2}\vert\vert C_{4z}]$ spin symmetry operation, which consists of a two-fold spin rotation combined with a fourfold lattice rotational symmetry. This symmetry leads to an altermagnetic $d_{x^{2}-y^{2}}$-wave type momentum-dependent spin-splitting in the band structure \cite{Smejkal2022Sep,Smejkal2022Dec}.

Our DFT bandstructure calculations (see Methods), shown in Fig.~\ref{fig:DFT}a confirm the presence of the $d_{x^{2}-y^{2}}$-wave type momentum-dependent spin-splitting and reveal a large magnitude reaching 1.5 eV in the valence band. The spin-orbit coupling is included in Fig.~\ref{fig:DFT}(a) and we see that it represents only weak corrections to the nonrelativistic altermagnetic band structure (see supplemental material (SM)~\cite{SM}. Consistently with previous reports \cite{Ablimit2018,Ma2021,Wu2024}, we obtain a metallic electronic structure of  V$_2$Te$_2$O monolayer as shown in the density of state calculations in Fig.~ \ref{fig:DFT}b.  

The total density of states (DOS) for the  V$_2$Te$_2$O monolayers also reveals distinct electronic singularities  near the Fermi level. Notably, a peak associated with a Van Hove singularity (VHS) appears approximately 0.7 eV above the Fermi level and around -1.4 eV below it. These VHS features can be tuned via electronic (hole) doping. 
Additionally, 
weaker DOS singularities are also observed directly near the Fermi level. 

While we focus on V$_2$Te$_2$O, we remark, that  metallic altermagnetism  was experimentally indicated in a sister compound KV$_2$Se$_2$O\cite{Jiang2024Aug}. We also point out that the V$_2$Te$_2$O lattice is conducive for a large AM spin splitting due to the Te and O atoms generating strong AM sublattice anisotropies which are around the two V sublattices related by a 90 degrees rotation. There exist, in principle, also 2D altermagnets with a smaller AM spin splitting. In the SM \cite{SM}, we show  the electronic  structure of CoS$_2$, where we observe much smaller AM spin splitting which is related to the weak rotations of the S polyhedra in opposite sense around the two Co atoms in the unit cell. The spin group symmetry analysis predicts that the momentum-dependent spin splitting in CoS$_2$ is of $d_{xy}$-type.


\section{Effective model for a 2D AM metal}
\label{sec:effmodel}

To capture the essential features of the unconventional spin splitting of the material candidates in Sec.~\ref{sec:Materials}, we set up an effective, minimal tight-binding model on the square lattice. 
The corresponding Hamiltonian is given by 
\begin{equation}
H_0=\sum_{\mathbf{k},\sigma,\sigma'}
\xi_{\mathbf{k}\sigma}c^{\dagger}_{\mathbf{k}\sigma}c_{\mathbf{k}\sigma}\,,
\label{eqn:H_0}
\end{equation}
where $c^\dagger_{\mathbf{k}\sigma}$ ($c_{\mathbf{k}\sigma}$) is the Fourier transform of the operator $c^\dagger_{i\sigma}$ ($c_{i\sigma}$), creating (annihilating) an electron with spin projection $\sigma=\pm$ for $\uparrow,\downarrow$ at lattice site $i$. Because the system is inside the AM phase, the 
dispersion $\xi_\mathbf{k\sigma}$ is modified by a momentum- and spin-dependent splitting 
\begin{align}
    \xi_{\mathbf{k}\sigma}&=\xi^0_{\mathbf{k}} - \lambda \sigma f(\mathbf{k})\,, \\
    \xi^0_{\mathbf{k}}&=-2t(\cos{k_x}+\cos{k_y})  -4t^{\prime}\cos{k_x}\cos{k_y}-\mu
    \,.
    \label{eqn:xi}
\end{align}
Here, $t$ labels 
nearest- and $t^\prime$ second-nearest-neighbor hopping amplitudes, 
$\mu$ the chemical potential, and $\lambda$ the magnitude of the AM splitting. 
Based on possible symmetry configurations \cite{Smejkal2022Dec}, we consider two types of AM form factors $f(\mathbf{k})$, which fall into the $B_1$ and $B_2$ irreducible representations of the $C_{4v}$ point group, described by the following basis functions
\begin{align} 
    f_{
    {x^2-y^2}}(\mathbf{k})&=\cos k_x -\cos k_y \label{eq: dx2y2 ff}\,,\\
    f_{
    {xy}}(\mathbf{k})&=\sin k_x\sin k_y \,. \label{eq: dxy ff} 
\end{align}

The material candidate $\mathrm{V}_2\mathrm{Te}_2\mathrm{O}$ exhibits an AM splitting described by $f_{x^2-y^2}$. We can  
approximate 
 its bands around the Fermi level 
by the dispersion Eq.~\eqref{eqn:xi} using the parameters $t=2\lambda/3,\,t^\prime=\lambda/5$. 
We show the minimal-model dispersion 
in Fig.~\ref{fig:DFT}c 
alongside the DFT band structure. 

The effective model for quasi-2D AM metals Eqs.~\eqref{eqn:xi}-\eqref{eq: dxy ff} is rooted in tetragonal symmetries. Thus, it can be expected to describe other AM metals with appropriately modified parameters. We take this as a motivation to analyze a comprehensive parameter regime beyond the specific material realizations. Our findings can guide additional material searches for further promising candidates. In general,
the single particle dispersion Eq.~\eqref{eqn:H_0}, exhibits two distinct behaviors in terms of the ratio   
$t/\lambda \lessgtr 1/2$ (see SM~\cite{SM}), which we dub small or large AM regime. They dictate where 
extrema and saddle points of the dispersion lie in the Brillouin zone (BZ). 
Both regimes are relevant for realistic material candidates as noted in Sec.\ref{sec:Materials}. 
The small AM regime $t/\lambda \gtr 1/2$ connects to the well-studied square-lattice dispersion for $\lambda=0$, where perfect nesting $\xi^0_{\mathbf{k+Q}}=-\xi^0_{\mathbf{k}}$ with $\mathbf{Q}=(\pi,\pi)$ occurs in combination with VH singularities at the Fermi level for $\mu=0,t'=0$. The VH points occur at the X $=(\pi,0)$ and Y $=(0,\pi)$ points of the BZ. 
When $t'$ is increased nesting is destroyed and the VH singularities move in energy according to
 $\mu^0_\mathrm{VH}=4t'$. 

\begin{figure}[t]
    \includegraphics[width=0.49\linewidth]{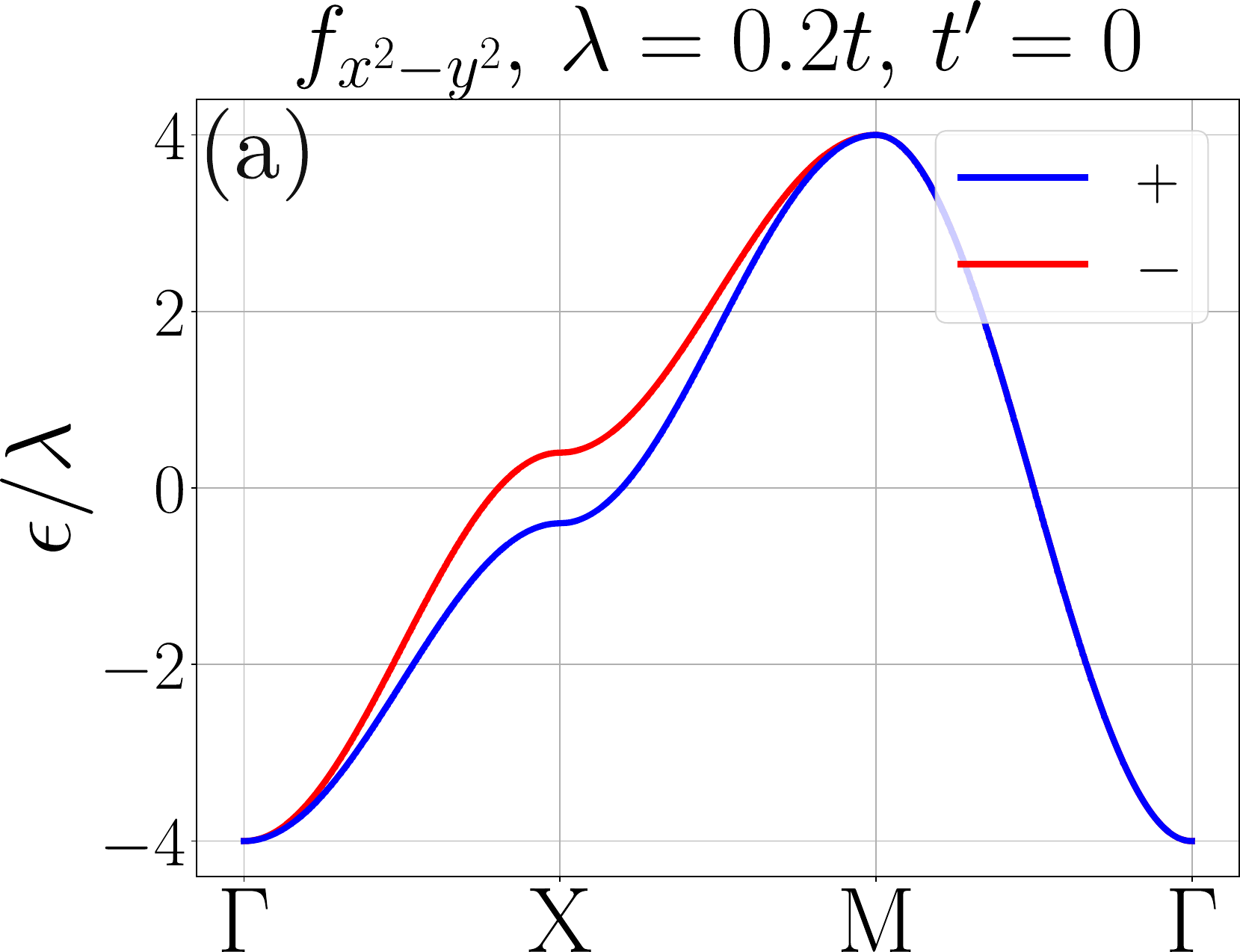}%
     \hfill\includegraphics[width=0.49\linewidth]{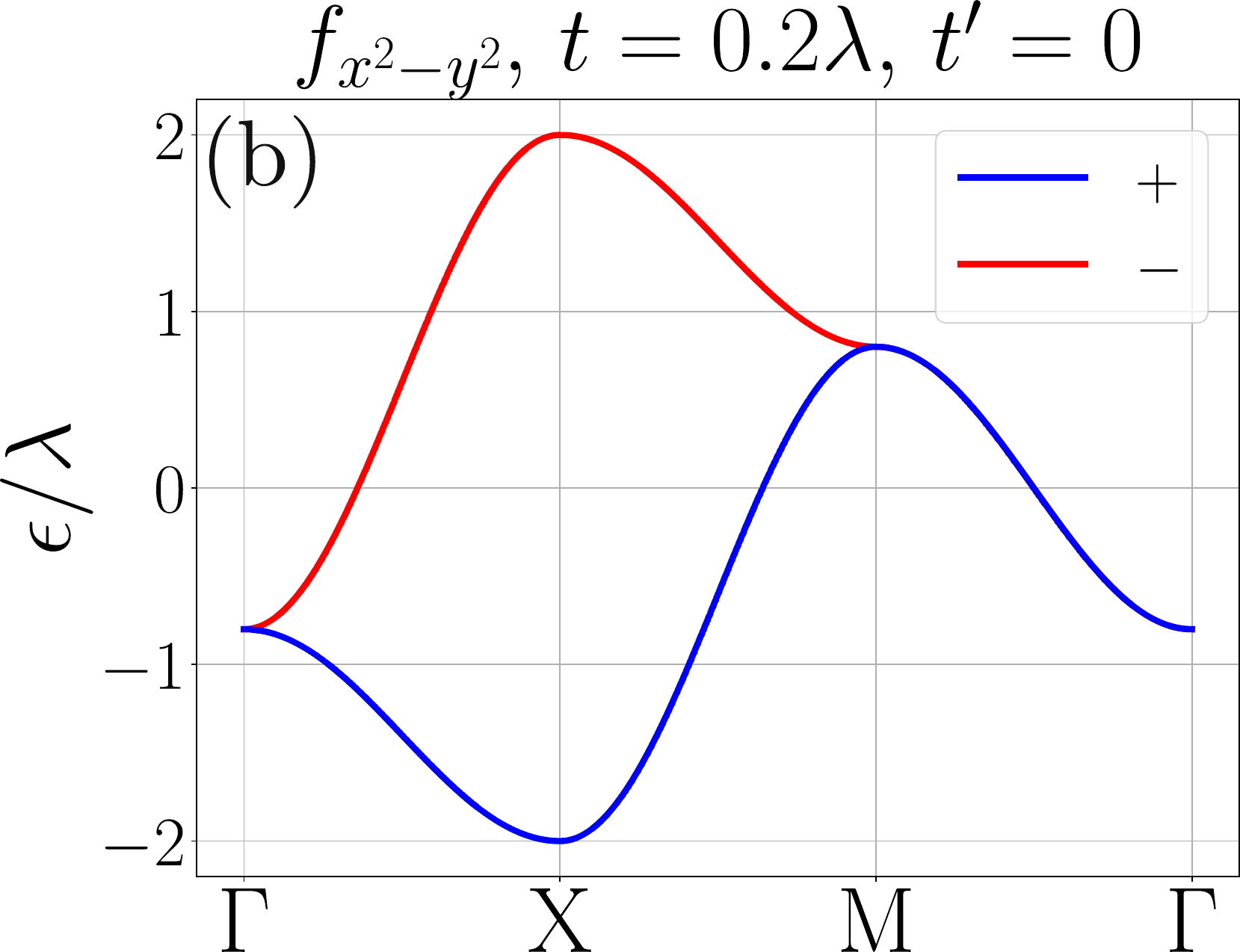}
     
     \includegraphics[width=0.49\linewidth]{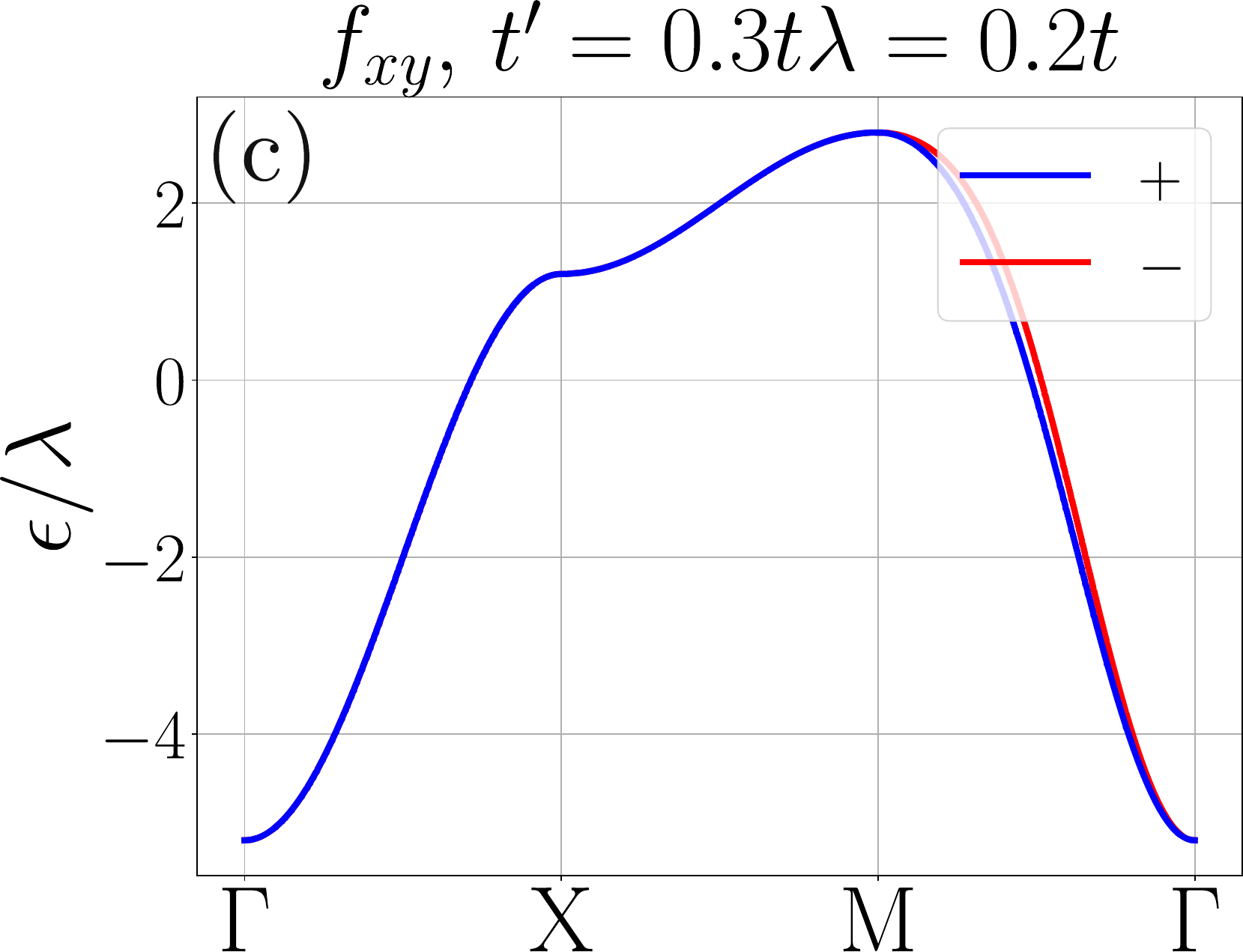}%
     \hfill\includegraphics[width=0.49\linewidth]{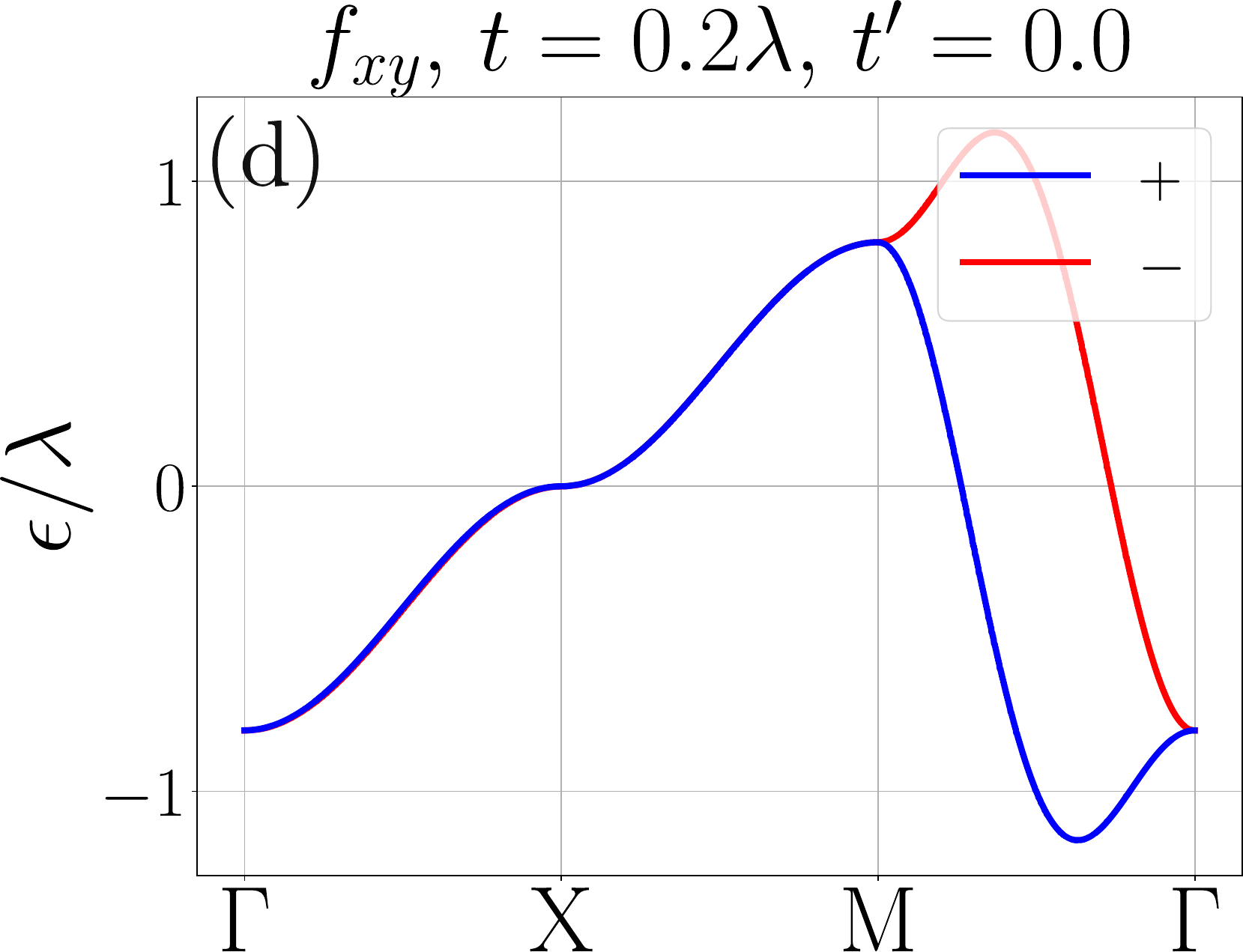}
\caption{\textbf{Effective models of an AM metal} according to the bare dispersion (Eq.(\ref{eqn:ham})) for the two AM form factors studied. The spin splitting is tuned by $\lambda$. The relative ratio  $t/\lambda$ tunes between the small (a,c) and large (b,d) altermagnetic regime and affects the location of saddle points and extrema in the BZ. } 
 \label{fig:bands}
\end{figure}

Adding $\lambda\neq0$ modifies this scenario differently for the two form factors $f_{x^2-y^2}$ and $f_{xy}$.
For  the $f_{x^2-y^2}$ symmetry of the AM splitting Eq.~\eqref{eq: dx2y2 ff}, nesting between equal spins remains intact $\xi_{\mathbf{k+Q}\sigma}=-\xi_{\mathbf{k}\sigma}$ at $\mu=0$, but is destroyed for opposite spins. The VH singularities 
are split in energy to chemical potential values $\mu_{\mathrm{VH}}=4t^\prime\pm2\lambda$, Fig.~\ref{fig:bands}a. The AM splitting is opposite at X and Y so that VH points become spin polarized at VH fillings. 
In the $f_{xy}$ case Eq.~\eqref{eq: dxy ff}, nesting is destroyed for equal spins but remains intact for opposite spins $\xi_{\mathbf{k+Q}\sigma}=-\xi_{\mathbf{k}-\sigma}$ at $\mu=0$. 
Spin-degenerate VH singularities still occur at X and Y.

In the large AM regime, when the dominant energy scale is the AM splitting $\lambda>2t$, 
the dispersion changes considerably. In the $f_{x^2-y^2}$ case, the minimum of the dispersion is found at X, while 
M $=(\pi,\pi)$ and $\Gamma=(0,0)$ points become spin-degenerate Van Hove singularities located at $\mu_\mathrm{VH}=\pm4t$, Fig.~\ref{fig:bands}b. 
In the $f_{xy}$ case, a VH singularity is still found at $\mu=0$  
and an additional occurs  for $\mu_{\mathrm{VH,\uparrow}}=4t$ 
at M and  and for $\mu_{\mathrm{VH,\downarrow}}=-4t$ at $\mathrm{\Gamma}$. Figs.\ref{fig:bands}(c-d).

We further equip the effective AM model with an 
interaction term $H=H_0+H_U$  
given by an on-site Hubbard repulsion
\begin{equation}
    H_U=U\sum_{i}n_{i\uparrow}n_{i\downarrow}\,,
    \label{eqn:H_U}
\end{equation}
where $U>0$, and $n_{i\sigma}=c^{\dagger}_{i\sigma}c_{i\sigma}$ labels the density operator. This form of (effective) onsite Hubbard repulsion of the altermagnetic quasi-particles is relevant for a strongly screened Coulomb repulsion.  We assume that renormalizations from high-energy modes are already included in the parameters of $H$, i.e., we consider not only the hopping amplitudes $t,t'$ and the AM splitting $\lambda$, but also $U$ as an effective parameter.

\section{Interaction induced instabilities}

Our aim is to understand the effects of correlations in the presence of altermagnetism. To this end, we study the different instabilities of an AM metal as a function of $\mu$, $t$, and $\lambda$ of the model given by Eqs.~\eqref{eqn:H_0} and \eqref{eqn:H_U}. Instabilities are signaled by singularities in the static, effective interaction, which can be defined as the amputated, one-particle irreducible (1PI) part of the two-particle Green's function $\Gamma_{\alpha\beta\gamma\delta}(\mathbf{k}_1,\mathbf{k}_2,\mathbf{k}_3,\mathbf{k}_4)=\langle c^\dagger_{\mathbf{k}_1\alpha}c^\dagger_{\mathbf{k}_2\beta}c_{\mathbf{k}_3\gamma}c_{\mathbf{k}_4\delta}\rangle_{1\mathrm{PI}}$. We employ random phase approximation (RPA) \cite{SM} and  
truncated-unity functional renormalization group (TUFRG)~\cite{Lichtenstein2017,Profe2022,Honerkamp2001Oct,Metzner2012,Platt2013Nov} to calculate $\Gamma$ and extract the critical temperature, type, and symmetry of an instability from its singular part, see Sec.~\ref{sec:Methods}. 
Within the TUFRG, fluctuation corrections of particle-particle and particle-hole type are summed up to infinite order on equal footing. Keeping just one of these channels amounts to an RPA or the Cooper ladder. Since the TUFRG takes their mutual feedback into account, it allows us to not only analyze RPA-type instabilities, but also the suppression or cooperation of different ordering channels. For example, this includes mechanisms to obtain charge orders or superconductivity from repulsive interactions \cite{Platt2013Nov,Metzner2012,Braun2024Jul}. We refer to 
the method section (Sec.~\ref{sec:Methods}) for more details on our TUFRG implementation.
For our calculations we use an intermediate value of the bare Hubbard $U$. Since both the altermagnetic form factor and the ratio $t/\lambda$ affect the bandwidth, we choose $U=3t$ for both $f_{x^2-y^2}$ and $f_{xy}$ in the small AM regime. For the large AM we choose $U=3\lambda$ for $f_{x^2-y^2}$ and $U=0.8\lambda$  for $f_{xy}$. All relevant quantities are given in units of the dominant energy scale $t$ ($\lambda$) in the small (large) AM regime. 

\begin{figure}[t]
    \includegraphics[width=0.8\linewidth]{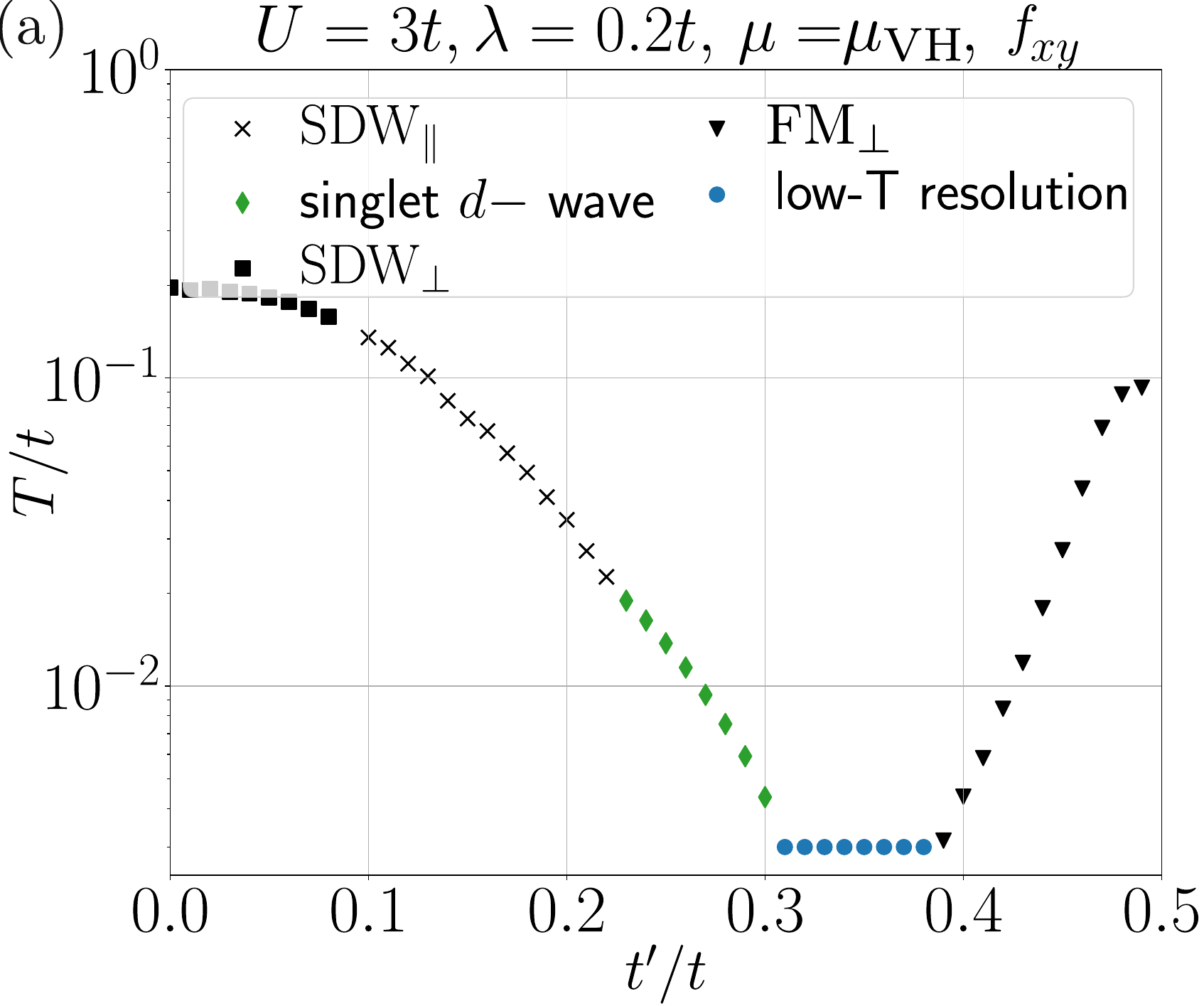}
     \includegraphics[width=0.8\linewidth]{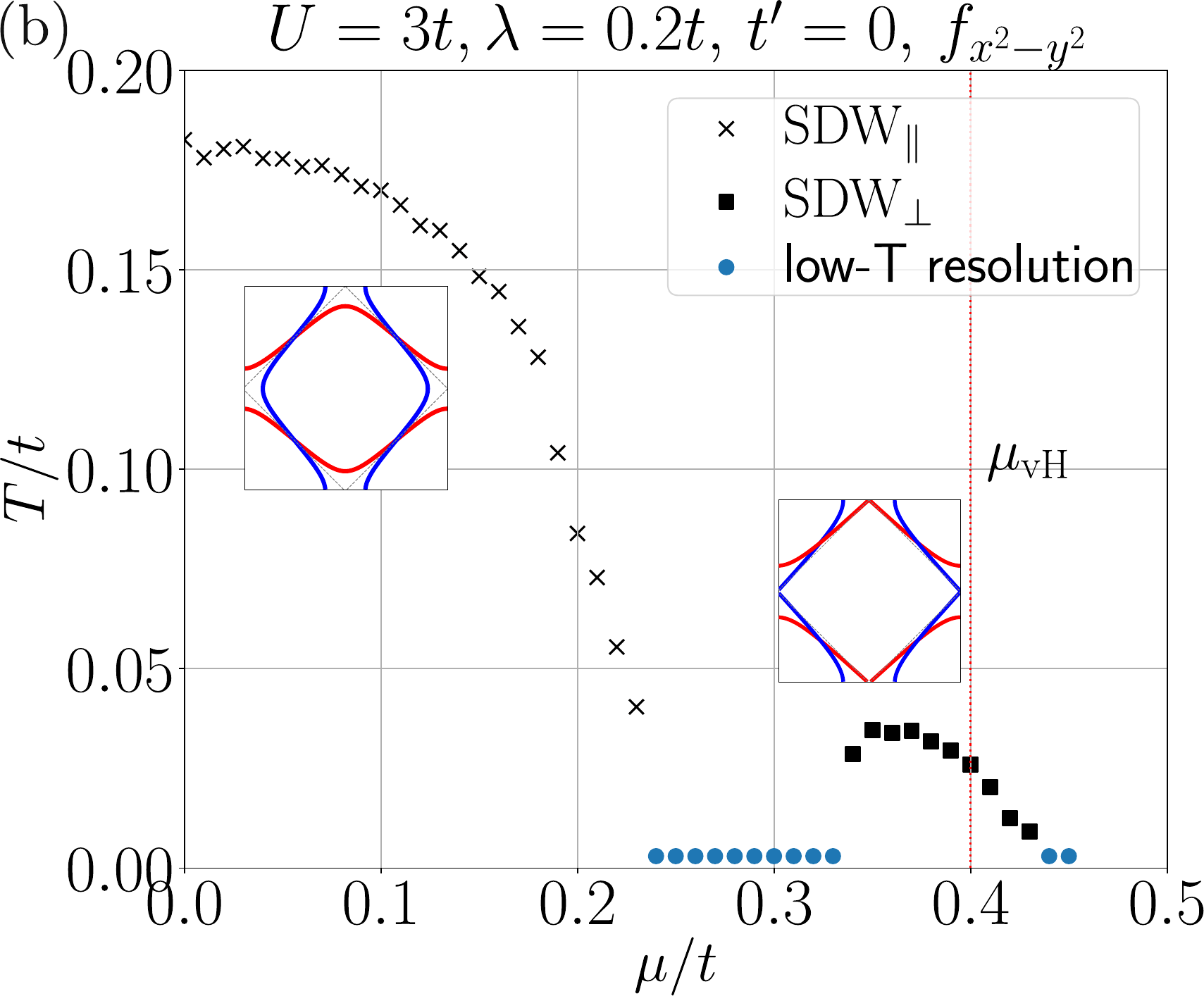}
\caption{\textbf{FRG phase diagrams for small the small AM regime.} (a) Critical temperatures $T$ of FRG instabilities as function of next-nearest-neighbor hopping $t'$ at $\mu_\mathrm{VH}=4t'$ for the $f_{xy}$ 
AM case. The results are qualitatively similar to the ones obtained in the SU(2) symmetric Hubbard model where for $t'\sim0.2t$ a transition from a 
SDW to d-wave pairing is observed. For larger values of $t'$ the system displays a ferromagnetic instability. In the $f_{xy}$ AM case, a transition between $\mathrm{SDW}{\perp}$ and $\mathrm{SDW}{\|}$ occurs. 
(b) For $f_{x^2-y^2}$ the observed instabilities $\mathrm{SDW}_{\|}$ and $\mathrm{SDW}_{\perp}$ are separated into two domes as a function of $\mu$ within our temperature resolution. The former is dominated by equal spin nesting at $\mu=0$ and the latter due to the spin split VH singularities occurring at $\mathbf{X}$ and $\mathbf{Y}$. The different symbols correspond to the instabilities identified with the vertex divergence terminating the FRG flow. The insets correspond to relevant FS for the instabilities identified.}
\label{fig:fRG_small_lambda}
\end{figure}

 \begin{figure*}[t]
        \centering
        \includegraphics[width=\linewidth]{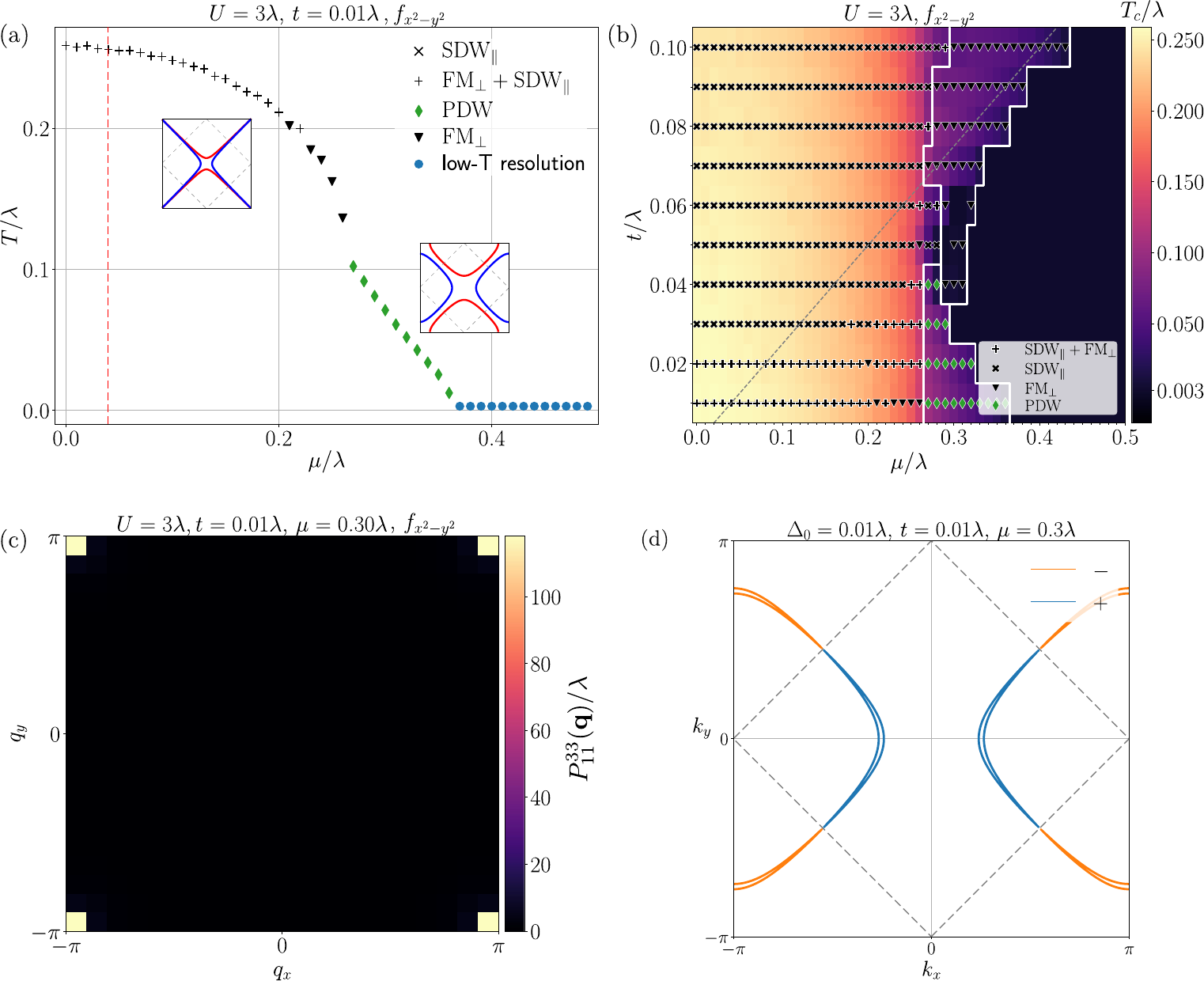}
        \caption{\textbf{FRG phase diagram and PDW in the large $f_{x^2-y^2}$ altermagnetic regime.} (a,b) 
        Critical temperature $T_c$ of FRG instabilities as function of chemical potential $\mu$ for fixed $t=0.01\lambda$ and varying $t$. We only show $\mu\geq0$ due to particle-hole symmetry. The dashed lines correspond $\mu_\mathrm{VH}=4t$. 
        The insets depict 
        example FSs for the observed instabilities. For  
        $t\lesssim0.04\lambda$ and for $\mu>0.27\lambda$ an instability towards a PDW arises. (c) The  
        singular part of the vertex for $t=0.01\lambda,\, \mu=0.3\lambda$ at the critical temperature. 
        The  
        singular contribution arises in the $m=0$ pairing vertex with extended $s$-wave symmetry $P_{11}^{33}(q)$
        and exhibits sharp peaks at $\mathbf{Q}=(\pi,\pi)$.  
        (d) The PDW Bogliubov quasiparticle bands indexed by $\pm$ exhibit a FS at $E=0$, for small values of the maximal gap magnitude $\Delta_0\lesssim t$. 
        For $\Delta_0\gtrsim t$, the Bogoliubov spectrum only
        possesses nodes 
        at $(k_x^{*},k_y^{*})=(\pm\arccos{\mu/2\lambda},\pm\arccos{\mu/2\lambda})$, 
        which is where the two quasiparticle bands touch due to the $s'$-wave form factor symmetry. These nodes are pinned at $E=0$ for any $\mu\leq2\lambda$. 
        }
        \label{fig:PDW}
    \end{figure*}

\paragraph*{Density waves.} 

The ordering tendencies we find can be understood as Fermi surface instabilities. It is therefore instructive to analyze the Fermi surface geometry. Two important ingredients in this regard are the presence of  
nesting and Van Hove singularities near the Fermi level in different parameter regimes of the effective model for an AM metal as detailed in Sec.~\ref{sec:effmodel}. When at least one of these features is present, RPA predicts an instability at sufficiently low temperatures. 
This is well known from the SU(2) symmetric Van Hove scenario on the square lattice, where a SDW appears as a result of nesting around Van Hove filling for $t'=0$ \cite{Honerkamp2001,Metzner2012}. In the AM case, we need to distinguish scattering processes of the same or opposite spins. Thus, we define the RPA susceptibility
\begin{equation}
    \hat{\chi}_{\perp/\|}(\mathbf{q})=\hat{\chi}^0_{\perp/\|}(\mathbf{q})\left[1-U\hat{\chi}^0_{\perp/\|}(\mathbf{q})\right]^{-1}\,,
    \label{eqn:RPA_def}
\end{equation}

with the bare ones given by
\begin{align}
   \chi_{\perp}(\mathbf{q})&=\frac{1}{2}\sum_{\mathbf{k},\sigma}\frac{n_{F}(\xi_{\mathbf{k}+\mathbf{q}\sigma})-n_{F}(\xi_{\mathbf{k}-\sigma})}{\xi_{\mathbf{k}+\mathbf{q}\sigma}-\xi_{\mathbf{k}-\sigma}}\,, \\
    \chi_{\|}(\mathbf{q})&=\frac{1}{2}\sum_{\mathbf{k},\sigma}\frac{n_{F}(\xi_{\mathbf{k}+\mathbf{q}\sigma})-n_{F}(\xi_{\mathbf{k}\sigma})}{\xi_{\mathbf{k}+\mathbf{q}\sigma}-\xi_{\mathbf{k}\sigma}}\,,
\end{align}
where 
$n_{F}(x)=\frac{1}{e^{\beta x}+1}$ is the Fermi function. Nesting or the presence of a VH singularity ensures that they will at least diverge $\propto\log T$ (see \cite{SM} for details). 
If either of them diverges at a finite $\mathbf{q}=\mathbf{Q}$ upon lowering temperature, they signal a 
density-wave instability, which we label as longitudinal ($\mathrm{SDW}_{\|}$) and transverse ($\mathrm{SDW}{\perp}$) spin density wave (SDW). They 
correspond to the order parameters 
\begin{align} 
\mathrm{SDW}_{\|}:\qquad \phi^z(\mathbf{Q})& =\sum_{\mathbf{k}
}\langle c^{\dagger}_\mathbf{k+Q}\sigma^z c_\mathbf{k}\rangle\,, \\
 \mathrm{SDW}{\perp}:\quad \phi^{x,y}(\mathbf{Q})&= \sum_{\mathbf{k}
}\langle c^{\dagger}_\mathbf{k+Q}\sigma^{x,y}c_\mathbf{k}\rangle\,, 
\end{align} 
with Pauli matrices $\sigma^a$ ($a=x,y,z$) and wave vector $\mathbf{Q}=(\pi,\pi)$ which can also turn incommensurate ($\mathbf{Q}=(\pm(\pi-\delta),\pi)$ or $\mathbf{Q}=(\pi,\pm(\pi-\delta))$, $\delta\ll 1$). These describe ordered states which spontaneously break translation symmetry, as well as the AM symmetry of simultaneous spin-flip and four-fold lattice rotation $[C_2\|C_4]$. As such the corresponding phase transition constitutes a novel type of symmetry breaking. The transverse SDW additionally breaks the remaining U(1) spin symmetry (while the longitudinal does not). 
\paragraph*{Competing orders in the small AM regime.} 

We start the discussion with the $f_{xy}$ form factor with $t^\prime=0$, as it provides a clear analogy to the 
well-studied Hubbard model on the square lattice ~\cite{Zanchi2000,Halboth2000,Halboth2000_PRL,Honerkamp2001,Honerkamp2001Oct,Husemann2009,Husemann2012,Metzner2012}. As noted before, a finite $\lambda$ breaks the spin SU(2) symmetry, but the VH singularity remains degenerate for the two spin species. 
Furthermore, there is only opposite spin nesting at $\mu=0$. 
This results in the antiferromagnetic order present in the SU(2) symmetric case becoming a $\mathrm{SDW}{\perp}$ for the same values of $\mu$ (Fig.~\ref{fig:fRG_small_lambda}a).

To investigate the effect of $t^\prime$ for finite $\lambda$
we look at the instabilities at $\mu=\mu_\mathrm{VH}$ as a function of $t^\prime$, which follow a similar trend to the SU(2) symmetric Hubbard model on the square lattice.  As $t^\prime$ is increased, the critical scale $T_c$  of the spin instabilities decreases because 
nesting is also destroyed. %
Additionally, there is a transition from $\mathrm{SDW}{\perp}$ to $\mathrm{SDW}{\|}$ because exactly at $t'=\lambda/4$ (keeping $\mu=\mu_\mathrm{VH}=4t'$) nesting between equal spins is restored promoting SDW$_{\|}$ as the dominant instability. 
After a threshold value ($t^\prime\sim0.2t$), the SDW$_{\|}$ instability gives way to a singlet pairing instability in the $d-$wave channel. This pairing instability arises despite the AM-induced mismatch $\xi_{\mathbf{k}\uparrow}\neq\xi_{-\mathbf{k}\downarrow}$, which, however, is zero at the VH points driving it. Further increase of $t^\prime$ leads to an upturn of $T_c$ in a ferromagnetically dominated regime along the transverse direction (FM$_\perp$). 
We can thus conclude that for the $f_{xy}$ form-factor the situation is almost identical to the well-known Van Hove scenario on the square lattice Hubbard model, with the only qualitative difference being in the magnetic instabilities that inherit the broken SU(2) symmetry of the single particle dispersion (Fig.~\ref{fig:fRG_small_lambda}a).    

We now turn to the $f_{x^2-y^2}$ form factor and $t'=0$. Our analysis yields a longitudinal commensurate SDW with $\mathbf{Q}=(\pi,\pi)$ close to $\mu=0$, where a condition of perfect equal-spin nesting is satisfied. This is signaled by diverging peaks at $\mathbf{Q}$ in the longitudinal particle-hole vertex. 
The longitudinal SDW order can become incommensurate as the chemical potential is tuned away from $\mu=0$. In doing so, the critical temperature decreases and drops to zero within our numerical resolution for $\mu\gtrsim 0.2t$.
Close to Van Hove filling $
\mu_\mathrm{VH}=\pm 2\lambda$, the AM metal becomes unstable to a transverse SDW. This can be understood by the fact that for this value of $\mu$, the spin up band has a Van Hove singularity only at the Y-point, while the spin down band has one only at X, as shown in the right inset in Fig.~\ref{fig:fRG_small_lambda}(b). Scattering processes between these two points in the BZ combined with spin flips favor a transverse SDW with $\mathbf{Q}=(\pi,\pi)$. The two instabilities are thus separated into two domes: one around $\mu=0$, dominated by $\mathrm{SDW}
_{\|}$, and one around $\mu_\mathrm{VH}$, dominated by $\mathrm{SDW}_{\perp}$. This effect becomes more pronounced as $\lambda$ is tuned to larger values.
Based on the previous discussion, we expect that the material candidate $\mathrm{V}_2\mathrm{Te}_2\mathrm{O}$, would fall into the small AM regime if residual interactions in the bands considered by our model are of the Hubbard type. 
(see SM\cite{SM}).
No superconducting instability is identified within our temperature resolution. Physically, this can be understood as the suppression of singlet pairing, 
in particular at the VH points due to the AM splitting, which violates the condition $\xi_{\mathbf{k}\uparrow}=\xi_{-\mathbf{k}\downarrow}$, thereby suppressing the so-called "Cooper logarithm" mechanism. 
We note, however, equal spin pairing states cannot be excluded on the basis of the above argument, 
and we cannot exclude that they occur at temperatures lower than the lowest value we can reach in our numerical calculations.

\paragraph*{Large AM regime and pair density wave.}

We now turn to the regime $t/\lambda < 1/2$, where the AM splitting sets the energy scale. 

In the $f_{xy}$ case for $U=0.8\lambda$ and small values of 
$\mu\lesssim0.05\lambda$, we find that the instability landscape as function of $\mu$ is dominated by the SDW$_\perp$ as $t$ is tuned to high values due to opposite spin nesting for both finite $t$ and $\lambda$ occurring at $\mu=0$. 
In a small region in the phase diagram, we also note the occurrence of instabilities  towards a longitudinal stripe order, where the divergent channel develops peaks at X and Y. This is due to the perfect equal-spin nesting in the particle-hole channel in the limit of $t=0$, for $\mathbf{Q}=\mathbf{X}$ or $\mathbf{Q}=\mathbf{Y}$.

For the $f_{x^2-y^2}$ form factor and small values of $t$, the system displays a strong competition between a transverse ferromagnetic (FM$_\perp$) and a $\mathbf{Q}=M$ SDW$_\|$ instability for several values of the chemical potential. In this case, for $\mu=0$, the Fermi surfaces are given by two almost "flat" (they are perfectly flat if $t=0$) diamonds centered around the X- and Y-points and both containing a Van Hove singularity at the M-point (see inset of Fig.~\ref{fig:PDW}a). For small values of nearest-neighbor hopping $t\leq 0.04\lambda$ we find that the two channels are nearly degenerate at the level of our FRG analysis.  
This is the result of a relatively large coupling $U/W=3/4$ which facilitates a Stoner instability towards FM$_\perp$. Furthermore, the transverse ferromagnetic instability is boosted due to the interchannel feedback from SDW$_\|$ enabled by the FRG. For smaller values of $U/W$,
the combination of nesting with a Van Hove singularity 
strongly favors the formation of a longitudinal SDW over a transverse ferromagnetic instability, which, based on the logarithmic scaling of the corresponding susceptibility, is always subleading. We confirm 
this behavior for smaller values of $U$, when the effects of the double logarithmic scaling of the particle-hole bubble are allowed to set in. For $U=2\lambda$, we find that the dominant instability for a wide range of values of $\mu$ is indeed the longitudinal SDW (see SM~\cite{SM}). 
As the chemical potential is tuned further away from zero $\mu\gtrsim0.3\lambda$ and
for larger values of $t\gtrsim0.05\lambda$, where the VH singularity is found away from $\mu=0$ and perfect equal-spin nesting is no longer present, FM$_\perp$ becomes the dominant instability (Fig.~\ref{fig:PDW}b). 
 
Finally, we show that the large AM regime facilitates an exotic pairing state mediated by magnetic fluctuations. For $\mu\gtrsim0.3\lambda$ and $t\lesssim0.05\lambda$ (Fig.~\ref{fig:PDW}b) we observe an instability towards a pair density wave (PDW) in the $m=0$ triplet pairing channel 
with a 
commensurate wave vector $\mathbf{Q}=\mathbf{M}$ (Fig.~\ref{fig:PDW}c) 
and an extended $s$-wave form factor ($s^{\prime}$). 
Thus, we 
extract the order parameter as 
\begin{equation} 
\Delta_{\mathbf{Q}} \equiv \sum_{\mathbf{k} \in \mathrm{BZ}} \Delta_0f^s_{\mathbf{k}} \langle c_{\mathbf{k+Q}}  \,(i\sigma^y\sigma^z) c_{-\mathbf{k}} \rangle \, , 
\end{equation} 
with $f^{s^{'}}_{\mathbf{k}}=\cos k_x + \cos k_y$.

The unusual combination of extended $s$-wave symmetry with a triplet pair is a special case for the general exchange symmetry of such a PDW gap function. We find that for unconventional finite-$q$ pairing the gap function must satisfy
\begin{equation}
    \Delta_{\mathbf{Q}}^{\sigma\sigma'}(\mathbf{k})=\langle c_{\mathbf{k+Q}\sigma} c_{-\mathbf{k}\sigma'}\rangle = - \Delta_{\mathbf{Q}}^{\sigma'\sigma}(\mathbf{-k+Q})\,,
\end{equation}
due to fermion statistics, i.e., the gap is odd under the combination of spin exchange and momentum inversion plus shift $\mathbf{k}-\rightarrow \mathbf{k+Q}$.
Similar to the SDW states discussed above, the PDW also breaks the lattice translation symmetry, as well as the $[C_2\|C_4]$ imposed by the altermagnetic term in Eq. (\ref{eqn:H_0}).  
The unconventional PDW state is similar to $\eta$ pairing \cite{PhysRevLett.63.2144,PhysRevLett.122.077002,PhysRevLett.74.789,PhysRevB.102.085140,PhysRevB.71.012512,PhysRevB.102.165150} or $\pi$ triplet pairing \cite{PhysRevLett.104.216403,Maitra2001,PhysRevB.62.9083,murakami1998backward}. However, a crucial difference is the presence of the AM term in the parent Hamiltonian and the unconventional mechanism generating the PDW state. 

The Bogoliubov spectrum of the PDW
state contains 
nodal points at $(k_x^{*},k_x^{*})=(\arccos{(-\mu/2\lambda)},\arccos{(\mu/2\lambda)})$ and $C_4$ symmetry equivalent locations. Since their location depends on the ratio $\mu/2\lambda$, they are always present in the BZ so long as $-1\leq\mu/2\lambda\leq1$.  For $\Delta_\mathbf{Q}\gtrsim t$, these are the only nodal points in the excitation spectrum. However, for $\Delta_\mathbf{Q}\lesssim t$, Bogoliubov Fermi surfaces arise (Fig.~\ref{fig:PDW}d).

The structure of the FS is again insightful in understanding the mechanism driving this particular instability. For $t=0$, there is perfect nesting in the pairing channel $\xi_{\mathbf{k+Q},\sigma}=\xi_{-\mathbf{k},-\sigma}$ for any $\mu$. For the values of $\mu$ where the dominant instability is towards the PDW, 
the FS occur in spin-split pockets centered around the X,Y points (inset Fig.~\ref{fig:PDW}a), which are connected by the high-symmetry wavevector $\mathbf{Q}=\mathbf{M}$. 
This means if the electron repulsion is turned attractive, there will be an instability in the Cooper channel for center-of-mass momentum $q=Q$.
For a finite $t$, $\xi_{\mathbf{k}+\mathbf{Q}\sigma}=\xi_{-\mathbf{k}-\sigma}$ no longer holds. However we expect for small enough values of $t$ for this effect to still persist. Indeed as can be seen in Fig.~\ref{fig:PDW}d, the PDW still occurs until $t\sim0.04$. 
Analyzing the inter-channel feedback within FRG, we find the main mechanism for the attraction to arise from both, SDW$_\|$ and FM$_\perp$ fluctuations. While the FRG contains more effects, this can be already understood in a spin-fluctuation approach restricted to projecting the spin vertex into the pairing vertex (see SM\cite{SM} for how this arises within FRG).  Close to a magnetic instability, the effective interaction is of the form $V_{spin}=\sum_{a=1}^3 V_a(\mathbf{q})S_\mathbf{q}^a S_\mathbf{-q}^a$ with $S_\mathbf{q}^q=\sum_\mathbf{k} c_{\mathbf{k+q}}^\dagger c_{\mathbf{k}}$. This leads to the properly antisymmetrized pairing interaction in the $m=0$ triplet channel
\begin{align}
    V_{zz}^{pair}(k,-k+q,k',-k'+q)&=\frac{1}{4}\Big[ V_x(k-k') + V_y(k-k') \notag \\
    &- V_z(k-k') \Big] - (k\leftrightarrow -k+q)\,.\label{eq:Vzzpair}
\end{align}
Note the opposite sign of transverse ($x,y$) and longitudinal ($z$) spin fluctuations, while the expression reduces to the standard $[V(k-k')-V(-k-k')]/4$ in the triplet channel for zero-momentum pairing $q=0$ in the SU(2) symmetric case $V=V_x=V_y=V_z$. Using that for FM$_\perp$ fluctuations $V_x(q)=V_y(q)$ peaks around $q=0$, and for SDW$_\|$ fluctuations $V_z(q)$ peak around $q=Q$, the interaction in the PDW channel with angular momentum $l$ is given by
\begin{align}
    V_{zz}^{pair,l}&=\sum_{k,k'} f_l(k) f_l(k') V_{zz}^{pair}(k,-k+q,k',-k'+q)\\
    &\approx \sum_k f_l(\pm k)\left[ f_l(k+Q)-f_l(-k) \right]\leq 0\,,
\end{align}
where the upper sign is for the FM and the lower for the SDW case. This PDW vertex is only non-zero for even lattice harmonics with $f_l(k+Q)=-f_l(k)$. Apart from the unusual pairing symmetry, the mechanisms for inducing a pairing attraction $V_{zz}^{pair,l}\leq 0$ are analogous to the well known cases of spin-fluctuation mediated pairing. (Transverse) FM fluctuations induce an attraction in the triplet channel. Longitudinal SDW fluctuations need to flip an extra sign ($-V_z$ in Eq.~\eqref{eq:Vzzpair}) and do so because the SDW wave vector connects regions of opposite sign of the gap function $f^s_{\mathbf k+Q}=-f^s_{\mathbf k}$.
We note that other pairing symmetries for the PDW triplet, such as $d$-wave, are subleading (see SM\cite{SM}). Furthermore, a PDW singlet would only be mediated by SDW$_\|$ not by FM$_\perp$ fluctuations and thus suffers from a smaller attraction. Finally, in the case of large $f_{xy}$ AM, there is also nesting in the pairing channel $\xi_{\mathbf{k+Q},\sigma}=\xi_{-\mathbf{k},\sigma}$, now for equal spin. However, fluctuations of the SDW$_\perp$ in the $f_{xy}$ AM phases diagram do not mediate an attraction for PDW triplet pairing, which explains why it is absent in the FRG phase diagram.

\section{Discussion}

We studied the phase diagram of the 2D Hubbard model on the square lattice in the presence of a $d$-wave AM as a function of several model parameters within a TUFRG approach. 
Remarkably, the inclusion of the AM splitting introduces two distinct energy regimes corresponding to small and large AM spin splitting.

Both regimes are feasible in realistic material candidates. The large AM spin splitting in metals was indicated experimentally in 3D RuO$_2$ \cite{fedchenko2024observation} or CrSb \cite{Reimers2024Mar} and in quasi2d KV$_2$Se$_2$O  \cite{Jiang2024Aug}. We also note that the tendency to secondary spin density wave described in our work, could be also a starting point for understanding the experimental indication of the spin density wave in KV$_2$Se$_2$O  \cite{Jiang2024Aug}.
The small AM splitting regime discussed for CoS in SM\cite{SM} is also experimentally feasible as it can be achieved by various different means including functionalization of more conventional antiferromagnetic monolayers such as antiferromagnetic FeSe \cite{Mazin2023}.

In both regimes, we found that the system as a function of the chemical potential, displays instabilities towards distinct SDW states either along the direction imposed by the AM splitting or perpendicular to it. 
The SDW states break the background AM $[C_2||C_{4z}]$ symmetry of two-fold spin and four-fold spatial rotation. 
In addition, correlations can induce pairing instabilities akin to the already studied mechanism in the  
SU(2)-symmetric Hubbard model in the small AM regime. 

When the AM dominates over the hopping term, we observe that the system can become unstable to an unconventional triplet PDW with commensurate wave vector $\mathbf{Q}$. 
This finite-momentum pairing instability is mediated by spin fluctuations and boosted  
by approximate nesting in the pairing channel $\xi_{-\mathbf{k}+\mathbf{Q},-\sigma}=\xi_{\mathbf{k},\sigma}$ of the $f_{x^2-y^2}$-wave AM term. Thus, the pairing state is stronger than for other finite-momentum states that rely on small splittings of the FS \cite{Zhang2024Feb,Sim2024Jul,Hong2024Jul,Chakraborty2024} and a larger critical temperature can be expected. We also showed that for such an unconventional PDW, the antisymmetry of the gap function needs to be adapted to include a momentum shift making spin triplet (singlet), even (odd) angular harmonics possible for appropriate lattice harmonics. 

The secondary instabilities we find are induced by  the interaction in the AM parent state, which we have included in the form of a Hubbard $U$. 
We use it to model the renormalized interaction that remains after the system undergoes the AM transition. 
The specific choice of the value of $U$ 
depends on the concrete material to model. Its value affects non-universal quantities such as critical temperatures and phase boundaries with respect to chemical potential $\mu$ and the relative strength of AM splitting $\lambda/t$. However, the instabilities
we identified are rooted in universal mechanisms: the spin density wave instabilities arise due to Fermi surface nesting and/or the presence of Van Hove singularities in the single-particle dispersion. Similarly, the pairing instabilities rely on the Cooper instability either from (approximate) inversion symmetry $\xi_{\mathbf k,\sigma}\approx\xi_{\mathbf{k},\sigma}$ or from (approximate) nesting in the pairing channel $\xi_{\mathbf k,\sigma}\approx\xi_{-\mathbf{k+Q},-\sigma}$ with increased critical temperatures due to the vicinity to a VH singularity. As such, these instabilities 
survive at arbitrarily weak coupling. 
The effect of longer-range components in the Coulomb interaction 
is an interesting  topic for future studies.

Since  the SDWs break translational symmetry, they can be experimentally probed by spin-resolved scanning tunneling spectroscopy measurements (STM). Furthermore, their Fermi surface reconstruction can be observed in ARPES measurements. Unfortunately, the specific PDW we find cannot be seen in STM, as the superconducting gap only changes sign from one sublattice to the other, there is no spatial modulation. 
However, signatures for its nodal points or Bogoliubov Fermi surfaces can, e.g., be observed via specific heat, thermal conductivity, or NMR measurements, while its specific nodal structure can be visualised via ARPES. In contrast to a zero-momentum superconducting state, an anisotropy occurs in thermal transport due to the breaking of $[C_2||C_4]$.

Our analysis also provides several further directions to study the effects of correlations in altemagnetic systems. It would be interesting to explore how the inclusion of sublattices and the symmetries of the underlying lattice 
modify the results presented here. This is 
not only relevant for other material candidates \cite{Smejkal2022Dec}, 
but also for "meta-altermagnets"; a recent theoretical proposal suggested twisted van der Waals materials as a platform for realizing altermagnetism with a high degree of tunability \cite{Liu2024Apr}. Furthermore, our approach can be generalized to include other symmetries of AM splitting like $g$-, or $i$-wave, and odd-partial wave unconventional spin splitting such as $p$ or $f$-wave \cite{Hellenes2024Jul}.  

Generally, our results show that 
altermagnetic metals can become unstable towards interaction-induced secondary phases. Thus, they can 
serve as a platform for the realization of exotic many-body states enabled by the unconventional spin splitting of the single-particle bands. It will be particularly interesting to find materials in the large AM regime, which can facilitate the investigation of the commensurate PDW.

\section{Methods} \label{sec:Methods}
\paragraph{DFT.} 
The electronic structure calculations for V$_2$Se$_2$O and V$_2$Te$_2$O were performed using the Vienna Ab-Initio Simulation Package (VASP) \cite{Kresse1996Oct,Kresse1996Jul} with the Generalized Gradient Approximation (GGA) as formulated by Perdew, Burke, and Ernzerhof (PBE) \cite{Perdew1996May}. A plane-wave basis set with a 500 eV kinetic energy cutoff was employed to ensure result accuracy. To account for the strong electron correlations in the V-3$d$ orbitals, the effective Hubbard method was employed with a  $U=4.0~$eV, consistent with previous $ab$-$initio$ studies \cite{Xing2020Sep,Cheng2021Sep,Zhu2023Dec}. The Brillouin zone was sampled using a $\Gamma$-centered 13$\times$13$\times$1  $\mathbf{k}$-mesh in the Monkhorst-Pack special scheme \cite{Monkhorst1976Jun}.
\paragraph{TUFRG.}
To resolve competition between ordering tendencies and account for interaction-induced renormalization, we employ a 
truncated unity FRG approach (TUFRG)~\cite{Lichtenstein2017,Profe2022} using a temperature cutoff~\cite{Honerkamp2001Oct}. The key object of our computations is the so-called vertex function $V$, which is the one-particle irreducible (1PI), amputated part of the two-particle Green's function. It corresponds to the sum of all 1PI diagrams with two incoming and two outgoing legs~\cite{Salmhofer1999,Kopietz2010,Metzner2012}. We neglect all frequency dependencies in $V$~\cite{freq_deps}, and parametrize it as 
\begin{equation}
    V^{\sigma_1\sigma_2\sigma_3\sigma_4}(\mathbf{k}_1,\mathbf{k}_2,\mathbf{k}_3;T)\,, 
\end{equation}
where $T$ is the temperature and $\sigma_i,\mathbf{k_i}$ label the spins and momenta of the incoming and outgoing fermions, respectively. We drop the temperature dependence for brevity, unless otherwise required. The momentum dependence can be further reduced to a channel decomposition \cite{Husemann2012,Husemann2009,Lichtenstein2017,Bonetti2024May}
\begin{equation}
    \begin{aligned}
     V^{\sigma_1\sigma_2\sigma_3\sigma_4}(\mathbf{k}_1,\mathbf{k}_2,\mathbf{k}_3)&=U(\delta_{\sigma_1\sigma_3}\delta_{\sigma_2\sigma_4}-\delta_{\sigma_1\sigma_4}\delta_{\sigma_2\sigma_3})\\
     &+P_{\mathbf{k}_1\mathbf{k}_3}^{\sigma_1\sigma_2\sigma_3\sigma_4}(\mathbf{k}_1+\mathbf{k}_2)\\
     &+D_{\mathbf{k}_1\mathbf{k}_4}^{\sigma_1\sigma_2\sigma_3\sigma_4}(\mathbf{k}_3-\mathbf{k}_1)\\
     &-D_{\mathbf{k}_1\mathbf{k}_3}^{\sigma_1\sigma_2\sigma_4\sigma_3}(\mathbf{k}_2-\mathbf{k}_3)\,. 
    \end{aligned}
\end{equation}
Here, $P,D$ correspond to the particle-particle and particle-hole channel. Notice the momenta appearing both as indices and as arguments in the channels. The latter, often characterized as a "strong" dependence provide information about the wave vector at which the instability occurs. The former "weaker" dependence, we expand in form factors. Due to the broken SU(2) symmetry on the single-particle level, we also expand the channels in terms of Pauli matrices. This leads to
\begin{equation}
    \begin{aligned}
        P_{\mathbf{k}\mathbf{k}'}^{\sigma_1\sigma_2\sigma_3\sigma_4}(\mathbf{q})&=\frac{1}{2}\sum_{\alpha,\beta=0,}^3 \sum_{\ell,\ell'} P^{\alpha\beta}_{\ell\ell'}(\mathbf{q})t^{\alpha}_{\sigma_1\sigma_2}(t^{\beta}_{\sigma_3\sigma_4})^\dagger f^{\ell}_\mathbf{k}(f^{\ell'}_{\mathbf{k}'})^* \\
        D_{\mathbf{k}\mathbf{k}'}^{\sigma_1\sigma_2\sigma_3\sigma_4}(\mathbf{q})&=\frac{1}{2}\sum_{\alpha,\beta=0,}^3 \sum_{\ell,\ell'} P^{\alpha\beta}_{\ell\ell'}(\mathbf{q})\tau^{\alpha}_{\sigma_1\sigma_2}\tau^{\beta}_{\sigma_3\sigma_4} f^{\ell}_\mathbf{k}(f^{\ell'}_{\mathbf{k}'})^*\,,
    \end{aligned}
\end{equation}
where $\tau^\alpha$ label the Pauli matrices (including unity), $t^\alpha=i\tau^y \tau^\alpha$ and
$\{f^{\ell}_{\mathbf{k}}\}$ labels the set of form factors. They constitute a complete basis, i.e. 
\begin{equation}
    \sum_{\ell,\ell'}f^{\ell}_{\mathbf{k}}(f^{\ell'}_{\mathbf{k}'})^*=\delta_{\mathbf{k}\mathbf{k}'}\,.
\end{equation}
We choose $\{f^{\ell}_{\mathbf{k}}\}$ such that they transform under the irreducible representations of $C_{4v}$ and truncate  at the first shell of nearest neighbors for our flow equations. That is, we only consider the following basis functions
\begin{align}
    &f_{\mathbf{k}}^0=1\\
    &f_{\mathbf{k}}^1=\cos{k_x}+\cos{k_y}\\
    &f_{\mathbf{k}}^2=\cos{k_x}-\cos{k_y}\\
    &f_{\mathbf{k}}^3=\sqrt{2}\sin{k_x}\\
     &f_{\mathbf{k}}^4=\sqrt{2}\sin{k_y}\,.
\end{align}
Within the 1-loop truncation \cite{Metzner2012,Platt2013Nov,Honerkamp2001Oct} we get the following flow equations
\begin{equation} 
\begin{aligned}
    \Dot{X}^{\alpha\beta}_{\ell\ell'}(\mathbf{q})=
    &\frac{1}{2}\zeta_{X}\sum_{\alpha',\beta'}\sum_{m,m'}X[V]^{\alpha\beta'}_{\ell m'}(\mathbf{q})\Dot{\Pi}^{\beta'\alpha'}_{X,mm'}(\mathbf{q})X[V]^{\alpha'\beta}_{m'\ell}(\mathbf{q})\,,
\end{aligned}
\label{eqn:floweqns}
\end{equation}
where $X={P,D}$, $\zeta_{D}=2,\zeta_{P}=-1$ and $\Dot{X}=\frac{d}{dT}$. $X[V]$ defines the projection of the full vertex onto channel $X$ and reads
\begin{widetext}
\begin{align}
  P^{\alpha\beta}_{\ell\ell'}[V](\mathbf{q})&=
  \frac{1}{2}\int_{\mathbf{k}\mathbf{k}'}\sum_{\substack{\sigma_1,\sigma_2\\\sigma_3,\sigma_4}} V^{\sigma_1\sigma_2\sigma_3\sigma_4}(\mathbf{k},\mathbf{q}-\mathbf{k},\mathbf{k}')(t^\alpha)^\dagger_{\sigma_1\sigma_2}t^\beta_{\sigma_3\sigma_4}f^\ell_{\mathbf{k}}f^{\ell^\prime}_{\mathbf{k}'}\\
  D^{\alpha\beta}_{\ell\ell'}[V](\mathbf{q})&=\frac{1}{2}\int_{\mathbf{k}\mathbf{k}'}\sum_{\substack{\sigma_1,\sigma_2\\\sigma_3,\sigma_4}} V^{\sigma_1\sigma_2\sigma_3\sigma_4}(\mathbf{k},\mathbf{k}'+\mathbf{q},\mathbf{k})\tau^\alpha_{\sigma_1\sigma_2}\tau^\beta_{\sigma_3\sigma_4}f^\ell_{\mathbf{k}}f^{\ell^\prime}_{\mathbf{k}'}\,.
\end{align}
\end{widetext}
Furthermore, the fermion loop integrals are given by
\begin{align}
    \Pi^{\alpha\beta}_{D}(\mathbf{q})&=-\frac{1}{2\beta}\sum_{i\nu}\int_{\mathbf{k}}\mathrm{tr}[\tau^{\alpha}\,G(\mathbf{\mathbf{k}+\mathbf{q}},i\nu;T)\,\tau^{\beta}\,G(\mathbf{k},i\nu;T)] \\
     \Pi^{\alpha\beta}_{P}(\mathbf{q})&=-\frac{1}{2\beta}\sum_{i\nu}\int_{\mathbf{k}}\mathrm{tr}[t^{{\alpha}\dagger}\,G({\mathbf{q}-\mathbf{k}},-i\nu;T)\,t^{\beta}\,G(\mathbf{k},i\nu;T)]\,,
\end{align}
where $G$ labels the free propagator, renormalized by the temperature
\begin{equation}
    G(\mathbf{k},i\nu;T)=T^{1/2}(i\nu-\xi_{\mathbf{k}\sigma})^{-1}\,.
\end{equation}
We integrate the flow equations given by Eq.~(\ref{eqn:floweqns}), until a temperature $T_c$ where the divergence condition $\mathrm{max}\{X\}\geq 100$ is met (given in units of $t$ or $\lambda$ depending on the regime studied) or until the numerical integration of the bubbles becomes unstable, which for our calculations is $T_{s}=3\times10^{-3}(t \, \mathrm{or}\, \lambda)$.
\section*{Acknowledgments}
We thank Daniel Agterberg, Brian Møller Andersen, Steffen Bollmann, Hannes Braun, Morten Christensen, Andreas Kreisel, Silvia Neri, Aline Ramires, Peng Rao, Michael Scherer, and Demetrio Vilardi for useful discussions. 
LC was supported by a grant from the Simons Foundation SFI-MPS-NFS-00006741-11. PMB acknowledges support by the German National Academy of Sciences Leopoldina through Grant No. LPDS 2023-06 and funding from U.S. National Science Foundation grant No. DMR2245246.
LS acknowledges support from the ERC Starting Grant No. 101165122.

\newpage
\bibliography{bibliography}
\newpage
\include{supp.tex}

\end{document}

%% file: supp.tex
\setcounter{section}{0}
\setcounter{equation}{0}
\setcounter{figure}{0}
\setcounter{table}{0}
\setcounter{page}{1}
\widetext
\begin{center}
\textbf{\large{Supplementary Material}}
\end{center}
\section{Minimal Hubbard model with altermagnetic splitting}
\label{sec:model}
 Here we provide more details on the effective single particle dispersion from the main text. We reiterate here that we do not study the mechanism responsible for the formation of the altermagnetic ordering, but rather use it as input for our calculations. The dispersion reads
\begin{equation}
    \xi_{\mathbf{k}\sigma\sigma^\prime}=[-2t(\cos{k_x}+\cos{k_y}) + 4t^{\prime}\cos{k_x}\cos{k_y}-\mu]\delta_{\sigma\sigma^\prime} - f(\mathbf{k})\sigma^z_{\sigma\sigma^\prime} \label{eqn:ham}
\end{equation}
Where $t$ is the nearest neighbor hopping, $t^\prime$ next-nearest neighbor hopping, $\sigma^x$, $\sigma^y$, and $\sigma^z$ are the Pauli matrices, and $\lambda(\mathbf{k})$ is a function that captures the symmetry of the altermagnetic order. In this article we focus on the two irreducible representations of the $C_{4v}$ group:
\begin{align}
    f_{x^2-y^2}(\mathbf{k})&=\lambda\left(\cos{k_x}-\cos{k_y}\right) \\
    f_{xy}(\mathbf{k})&=\lambda\sin{k_x}\sin{k_y}
\end{align}
where $\lambda$ tunes the magnitude of the splitting of the two spin bands.
\begin{figure}[ht!]
    \includegraphics[width=0.3\linewidth]{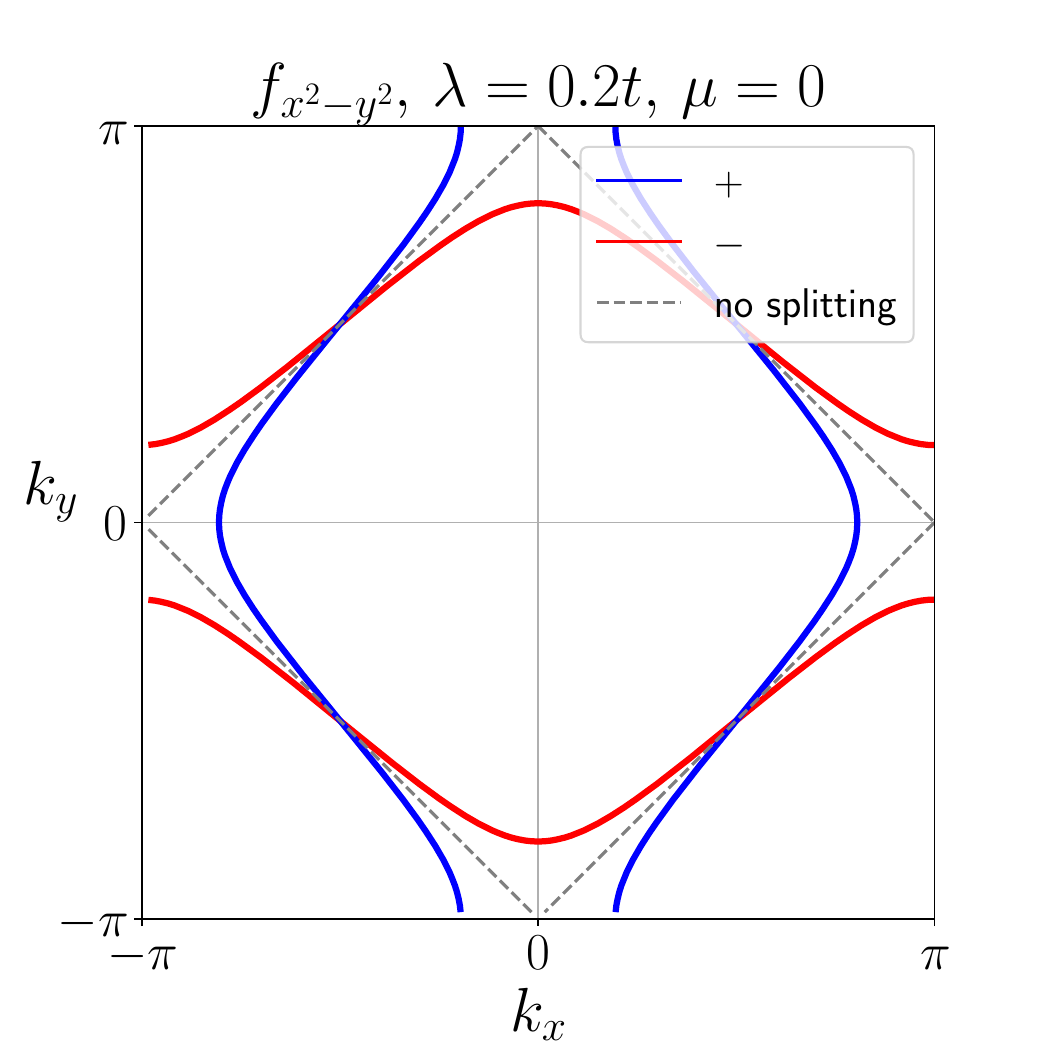}%
    \includegraphics[width=0.3\linewidth]{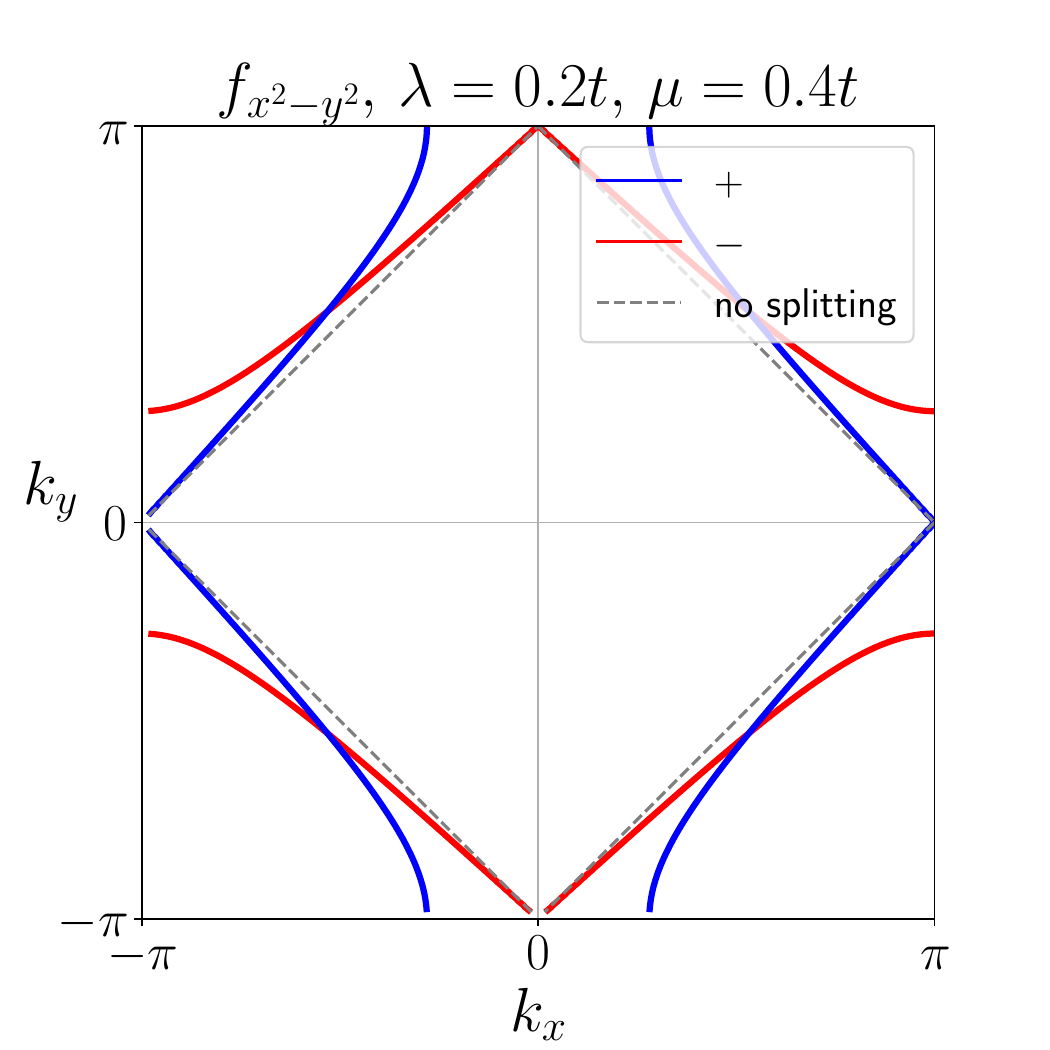}%
    \includegraphics[width=0.3\linewidth]{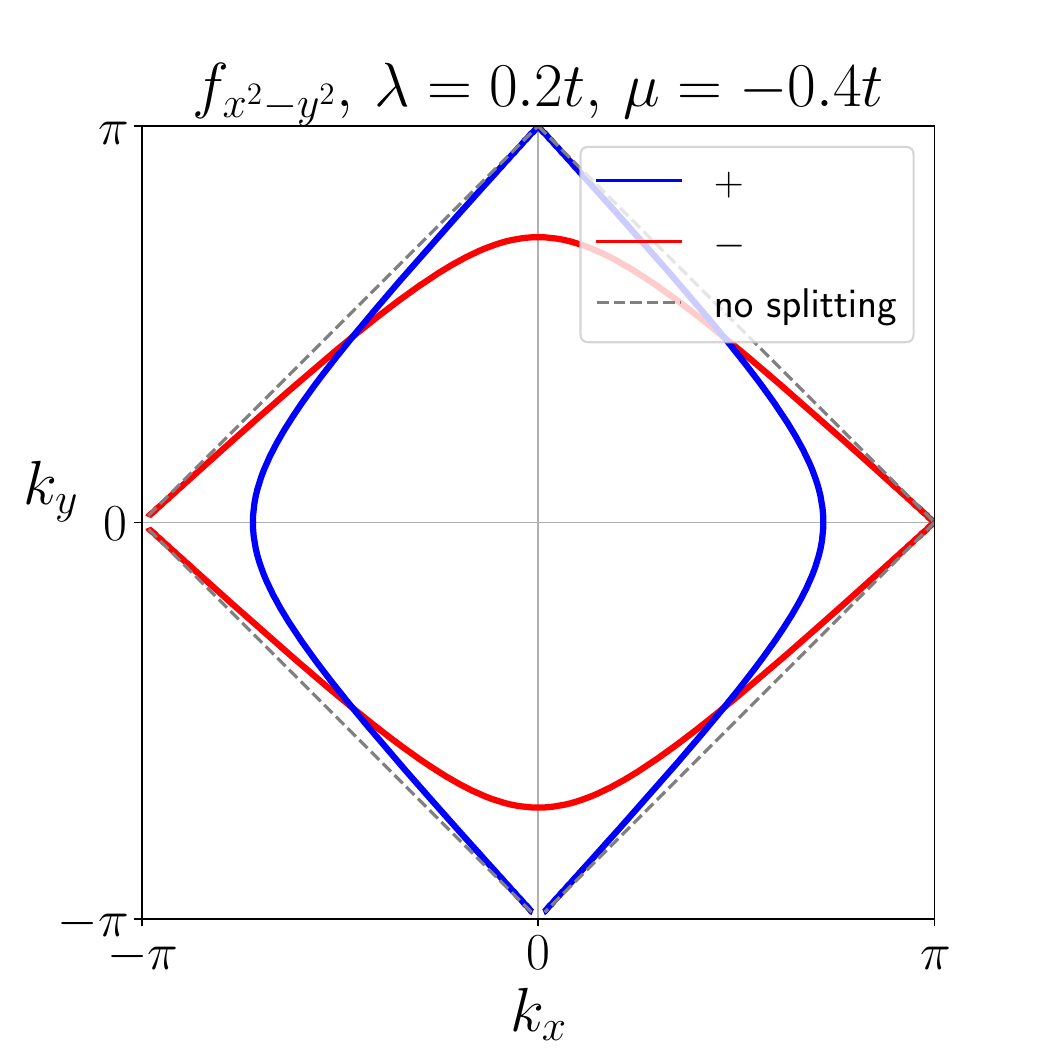} \\%
    \includegraphics[width=0.3\linewidth]{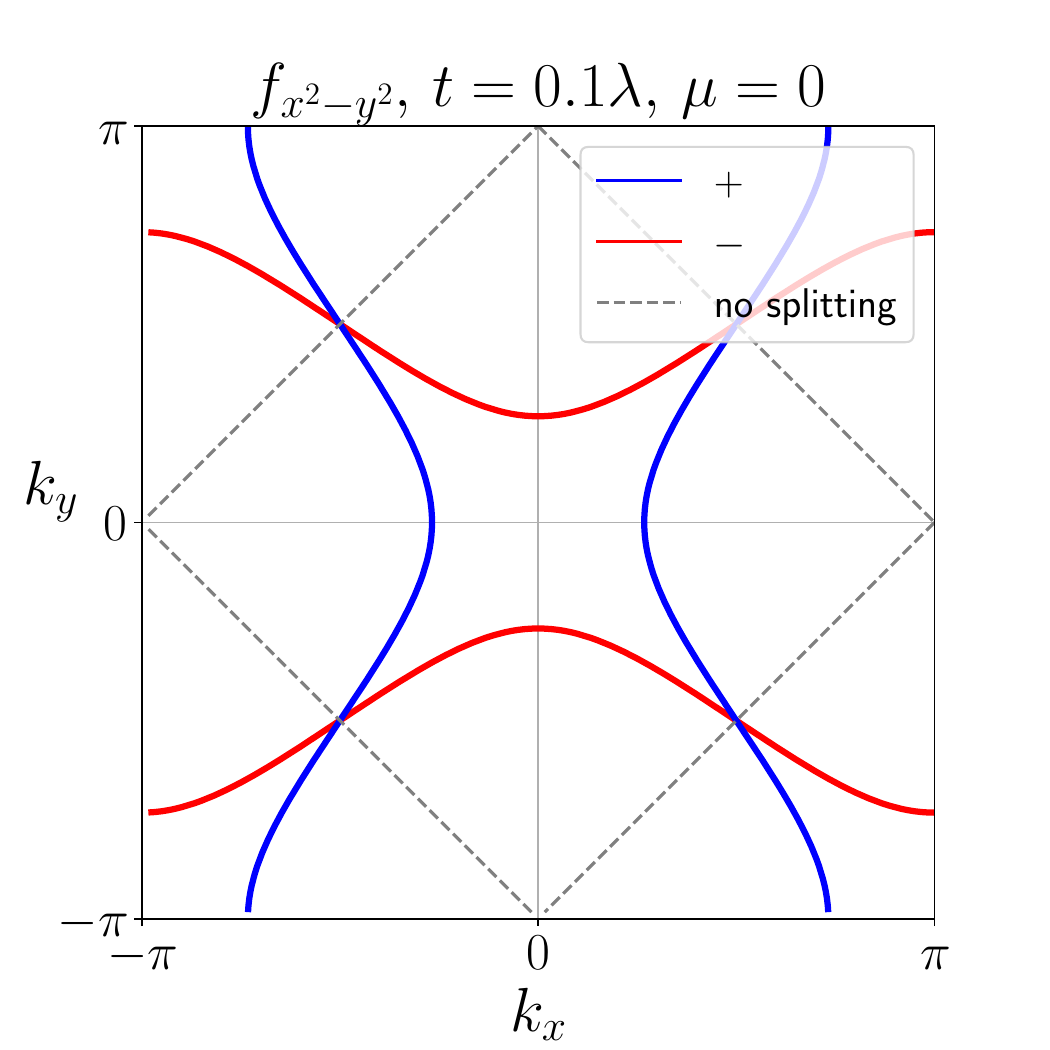}
    \includegraphics[width=0.3\linewidth]{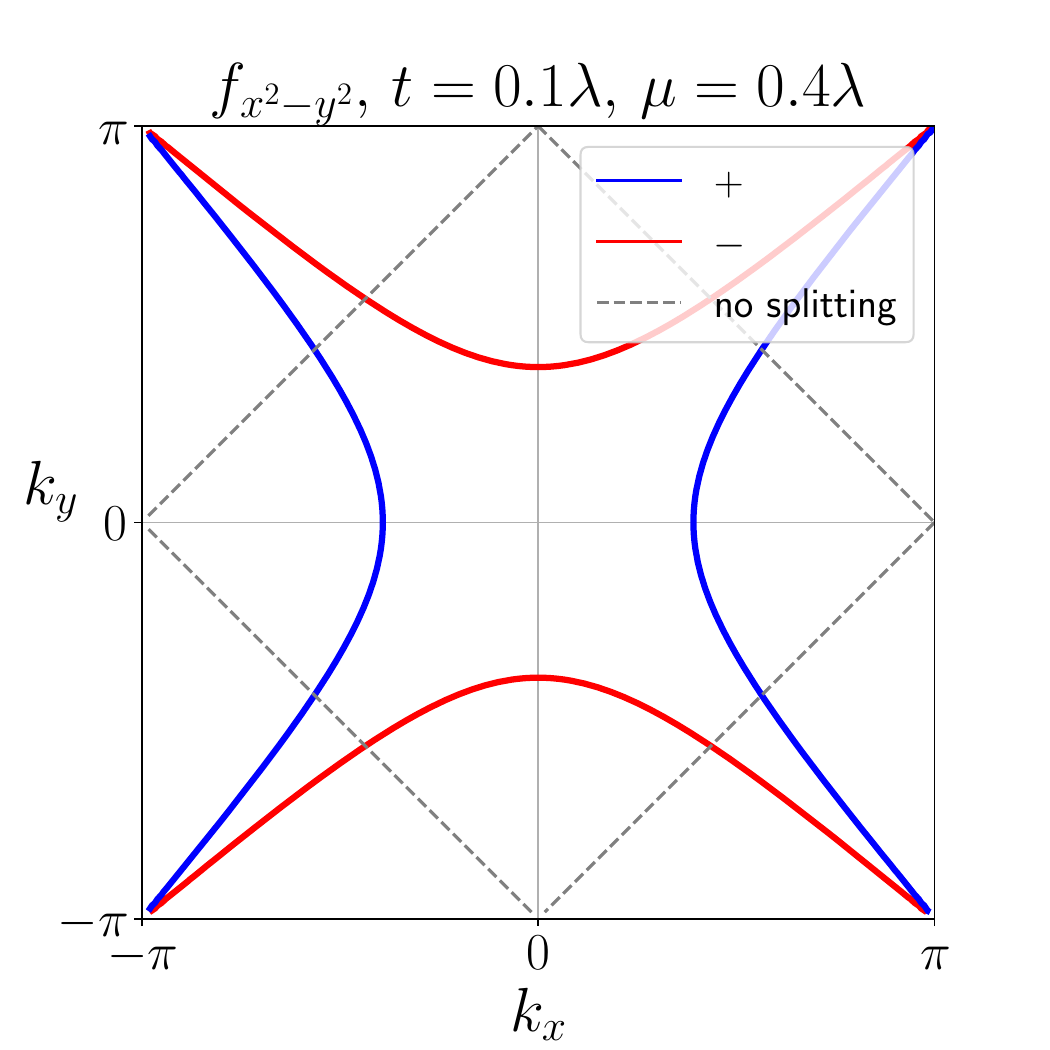}
    \includegraphics[width=0.3\linewidth]{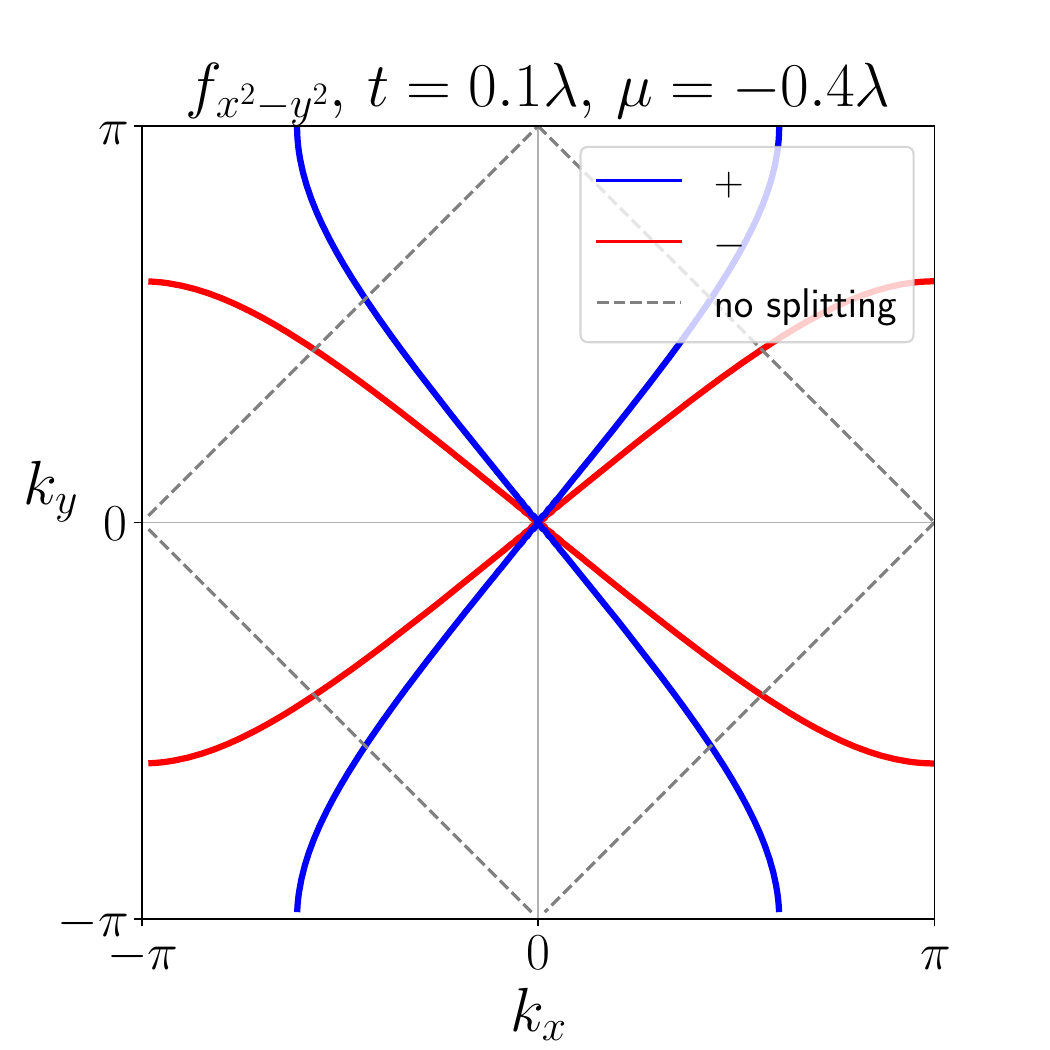}
    \caption{The Fermi surfaces for an $f_{x^2-y^2}$ altermagnetic metal for different chemical potentials $\mu$ and different ratios of hopping and AM splitting $t/\lambda$. For both the small and large AM regime, at $\mu=0$ there is perfect nesting for the same spin species. The location of the vH singularities as well as their spin degeneracy depends on the ratio $t/2\lambda$. In the small altermagnetic regime, they are spin split and are found at $\mathbf{X}$ and $\mathbf{Y}$. In the large AM regime they are found at $\mathbf{\Gamma}$ and $\mathbf{M}$.}
    \label{fig:FS_dx2-y2}
\end{figure}
It is instructive to study the features of the single particle dispersion for the two cases, in two distinct parameter regimes, namely $\frac{t}{\lambda}>\frac{1}{2}$ and $\frac{t}{\lambda}<\frac{1}{2}$. 
This arises from the Hessian determinant for $t'=0$
\begin{align}
    D_{x^2-y^2,\sigma}(\mathbf{k})&=(4 t^2 - \lambda^2) \cos k_x\cos k_y\,, \\
    D_{xy,\sigma}(\mathbf{k})&=-\lambda^2 \cos^2 k_x\cos^2 k_y +2t(\cos k_x -\sigma\lambda\sin k_x\sin k_y)(\cos k_y -\sigma\lambda\sin k_x\sin k_y)\,.
\end{align}
In both cases there is a critical value $t=\frac{\lambda}{2}$, which changes the extrema or saddle point character of the high-symmetry points of the Brillouin zone. This is crucial, as they dictate much of the landscape of instabilities that arise in this model, as we elucidate in the following sections.

We also note the presence of  perfect same spin nesting , $\xi_{\mathbf{k}+\mathbf{Q}\sigma}=-\xi_{\mathbf{k}\sigma}$ with $\mathbf{Q}=(\pi,\pi)$ is only present for the case of $f_{x^2-y^2}$, 
and perfect opposite spin nesting $\xi_{\mathbf{k}+\mathbf{Q}\sigma}=-\xi_{\mathbf{k}-\sigma}$ for the $f_{xy}$ form factor, which in both cases is destroyed by a finite $t^\prime$ and/or $\mu$.

The model Hamiltonian exhibits van Hove singularities, the details of which depend on the form of $\lambda(\mathbf{k})$ as well as the ratio $\frac{t}{\lambda}$. For $\frac{t}{\lambda}>\frac{1}{2}$, in the case of the $f_{x^2-y^2}$, the single particle dispersion is split for the two spins, along the $\Gamma$-$\mathrm{X}$-$\mathrm{M}$ direction, 

exhibiting two van Hove singularities at the X and Y points of the BZ, at chemical potential $\mu_{\mathrm{vH}}=4t^\prime\pm2\lambda$ (Fig. \ref{fig:FS_dx2-y2}). When the dominant energy scale is the altermagnetic splitting $\lambda$ which corresponds to the regime $t/\lambda<1/2$, the minimum of the dispersion is found at X and M becomes a vH singularity for spin up and a local maximum for spin down instead. Similarly, $\Gamma$ is a local minimum for spin down and a van Hove singularity for spin up (Fig.~\ref{fig:FS_dx2-y2}). 
\begin{figure}[t]
    \includegraphics[width=0.3\linewidth]{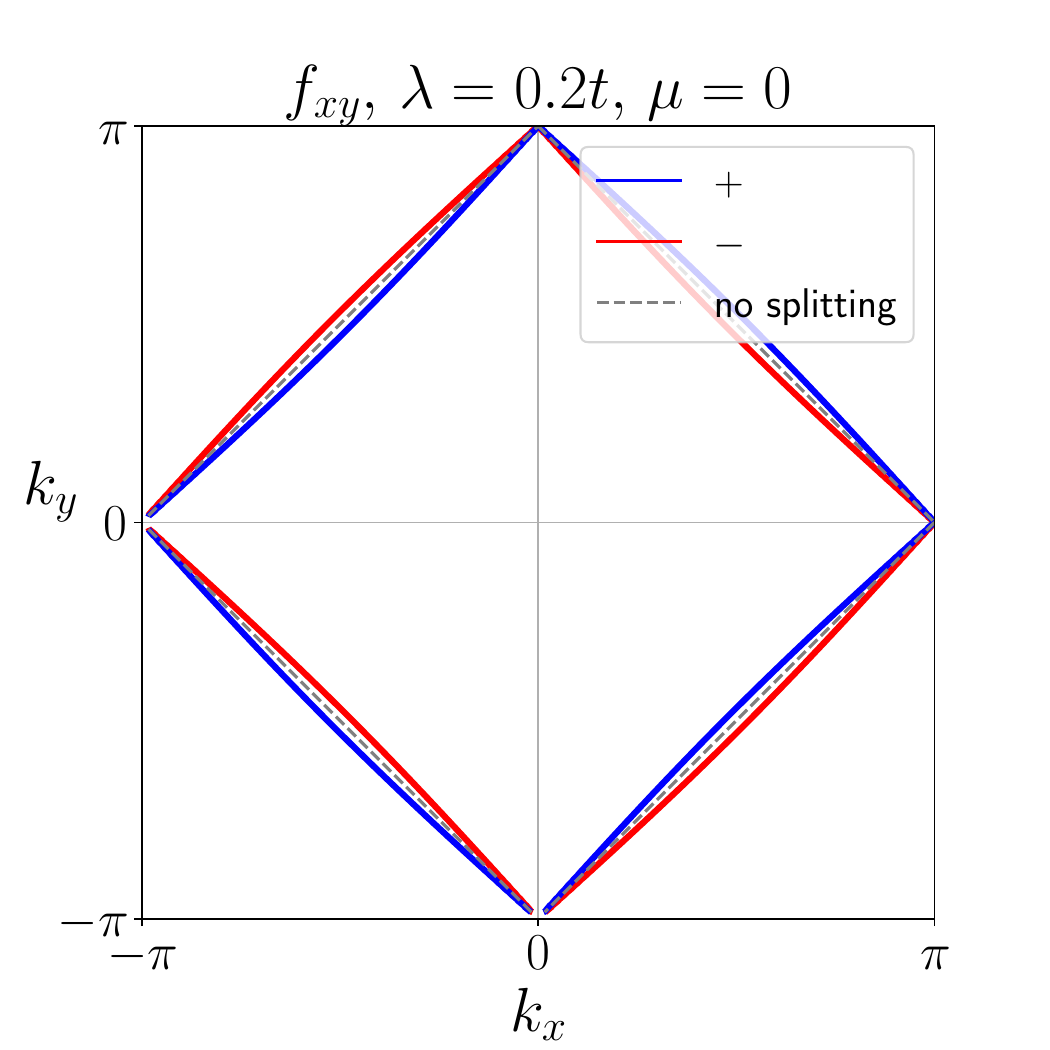}%
    \includegraphics[width=0.3\linewidth]{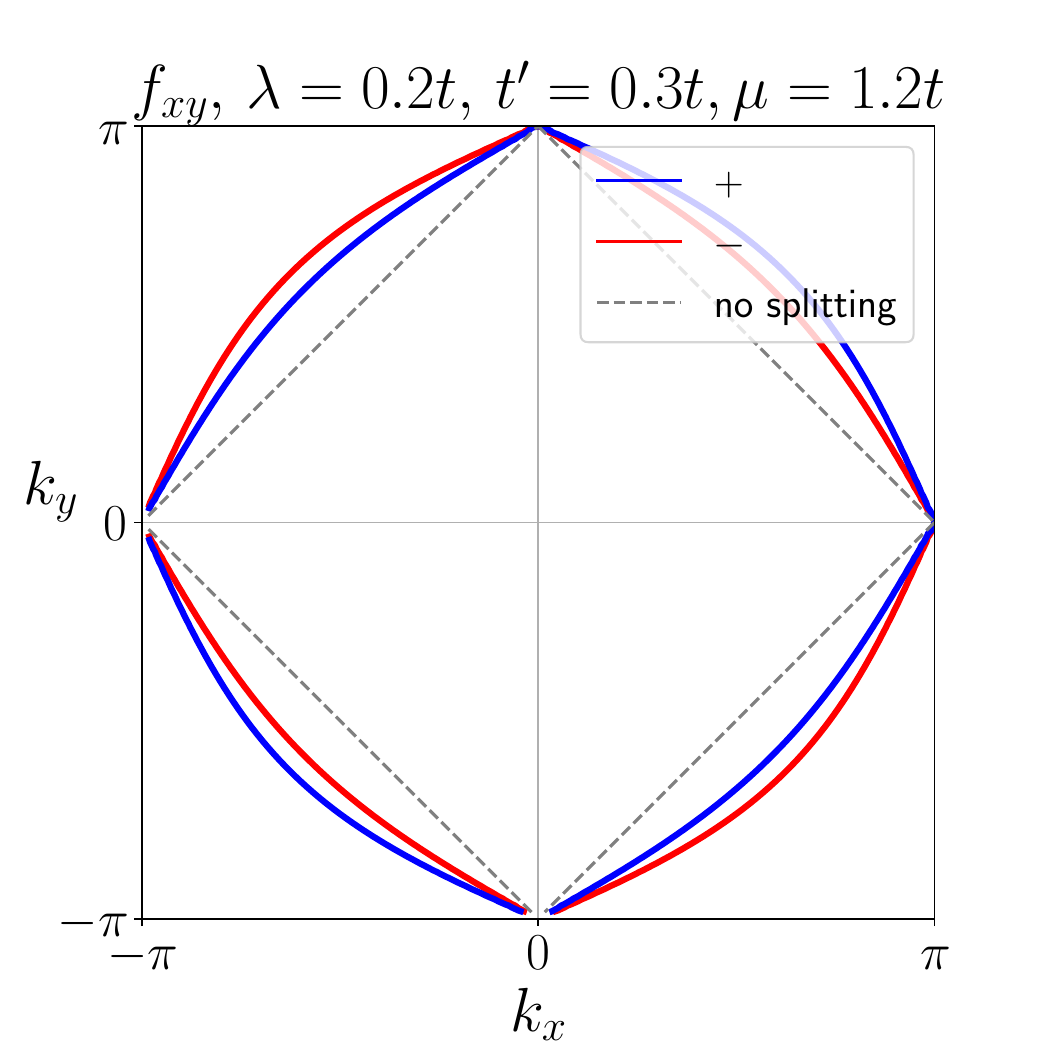}\\%
    \includegraphics[width=0.3\linewidth]{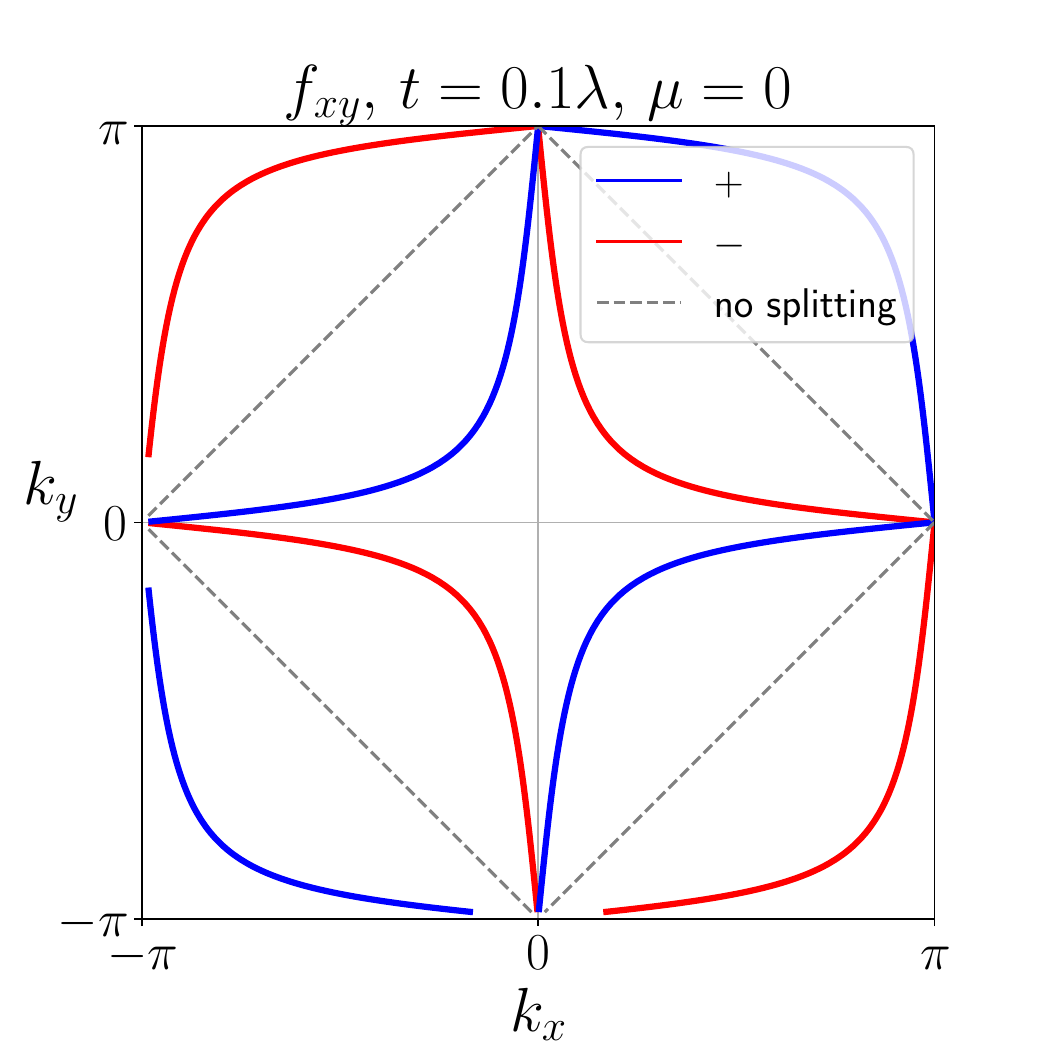}
    \includegraphics[width=0.3\linewidth]{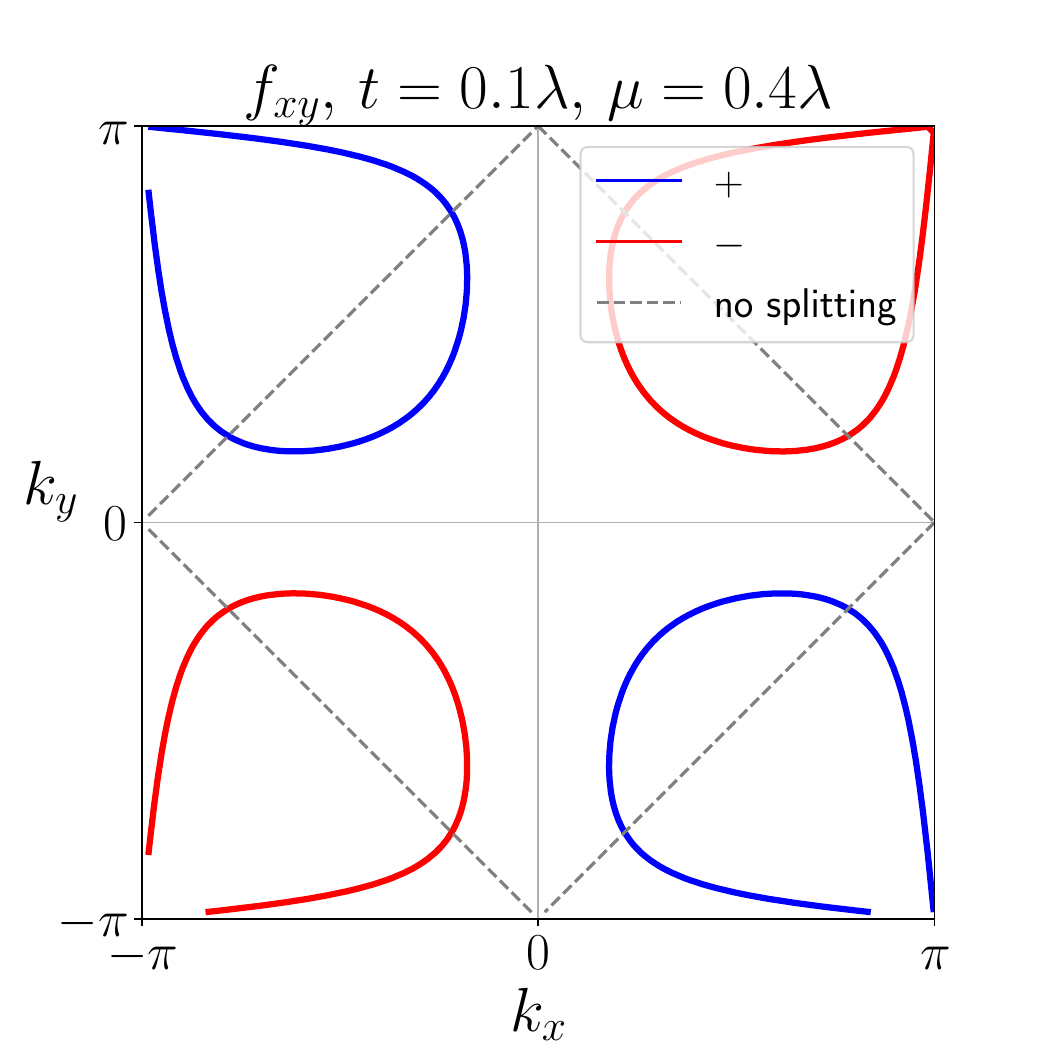}
    \includegraphics[width=0.3\linewidth]{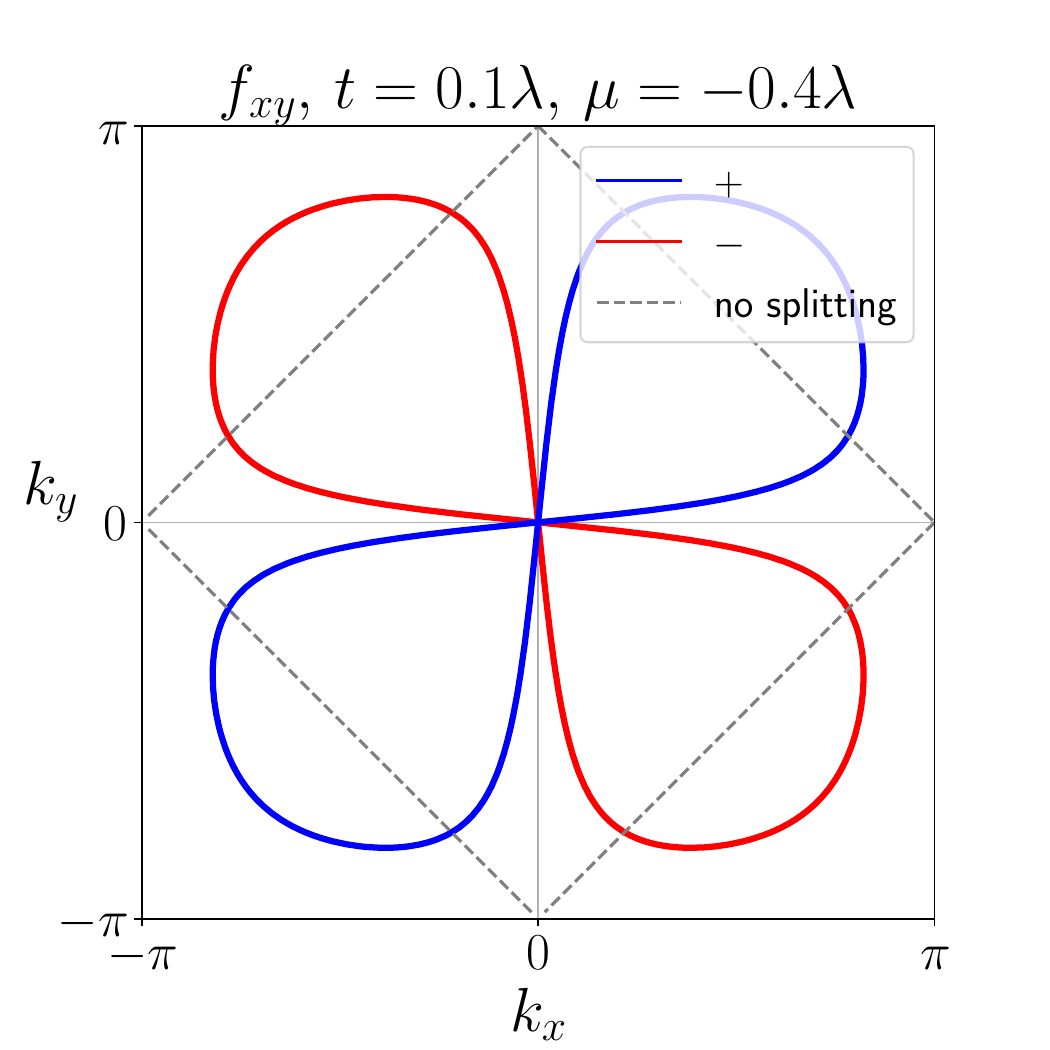}
    \caption{The Fermi surfaces for an $f_{xy}$ altermagnetic metal for different chemical potentials $\mu$ and different ratios of hopping and AM splitting $t/\lambda$. For this case both the small and large AM regimes posses perfect nesting for the opposite spin species at $\mu=0$.  Similar to the $f_{x^2-y^2}$ case, the location of the vH singularities as well as their spin degeneracy depends on the ratio $t/2\lambda$.}
    \label{fig:FS_dxy}
\end{figure}

For the $f_{xy}$ case, the splitting occurs along the $\mathrm{M}$-$\Gamma$ path with spin degenerate vH singularities at X and Y (\ref{fig:FS_dxy}). For $t/\lambda<1/2$, an additional van Hove singularity appears for spin up  for $\mu_{\mathrm{vH,\uparrow}}=4t$ at M at $\mathrm{\Gamma}$ for chemical potential $\mu_{\mathrm{vH,\downarrow}}=-4t$. The respective Fermi surfaces can be seen in Fig.~\ref{fig:FS_dxy} .
\section{Additional DFT band structure}
\begin{figure}
    \centering
    \includegraphics[width=0.5\linewidth]{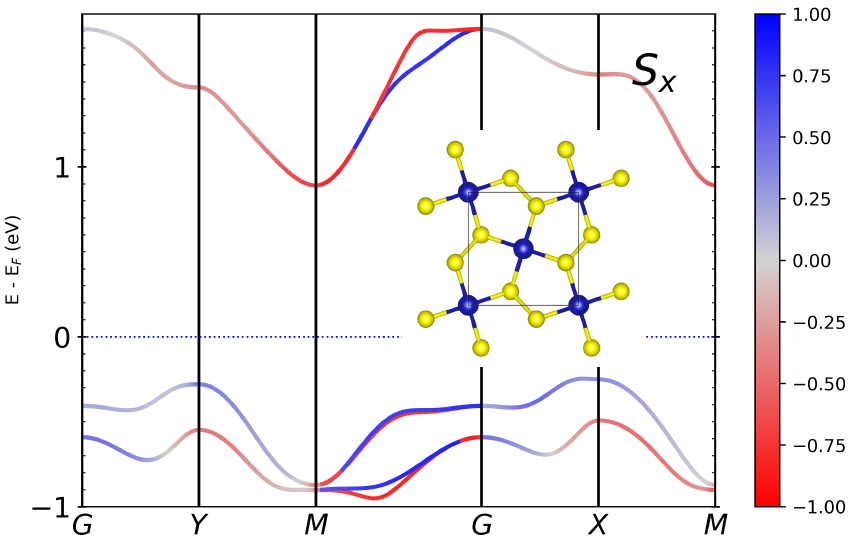}
    \caption{The DFT band structure for CoS$_2$ exhibits $d_{xy}$ altermagnetic splitting. Color corresponds to the projection of the spin along x-direction. The inset shows its lattice structure, with blue (yellow) corresponding to Co (S) atoms.}
    \label{fig:SM_COS2}
\end{figure}

We show the DFT band structure of CoS$_2$ in Fig.~\ref{fig:SM_COS2}. The AM splitting is related to weak rotation of S polyhedra in opposite sense around the two Co atoms in the unit cell .
\newpage
\section{Random Phase Approximation (RPA)}

\subsection{Temperature scaling of the bare spin susceptibility}

Evaluating the Matsubara sums in the static free susceptibility, otherwise known as the "particle-hole bubble", in the continuum limit we get 
\begin{align}
    \chi_{xx}(\mathbf{q})=\chi_{yy}(\mathbf{q})=\chi_{\perp}(\mathbf{q})&=\sum_{\sigma}\frac{(-1)^{\sigma}}{2}\int_{\mathbf{k}\in\mathrm{BZ}}\frac{n_{F}(\xi_{\mathbf{k}+\mathbf{q}\sigma})-n_{F}(\xi_{\mathbf{k}-\sigma})}{\xi_{\mathbf{k}+\mathbf{q}\sigma}-\xi_{\mathbf{k}-\sigma}} \\
    \chi_{zz}(\mathbf{q})=\chi_{\|}(\mathbf{q})&=\sum_{\sigma}\frac{1}{2}\int_{\mathbf{k}\in\mathrm{BZ}}\frac{n_{F}(\xi_{\mathbf{k}+\mathbf{q}\sigma})-n_{F}(\xi_{\mathbf{k}\sigma})}{\xi_{\mathbf{k}+\mathbf{q}\sigma}-\xi_{\mathbf{k}\sigma}}\,,
\end{align}
where $\sigma={+,-}$ labels the spin and $n_{F}(x)=\frac{1}{e^{\beta x}+1}$ is the Fermi function. These expressions can develop a logarithmic divergence if the nesting condition 
\begin{equation}
    \xi_{\mathbf{k}+\mathbf{q}\sigma}=-\xi_{\mathbf{k}\sigma^\prime}
\end{equation}
is satisfied for an appropriate combination of the spin indices $\sigma, \sigma^\prime$. If this condition is satisfied in the entirety of the FS ($\xi_{\mathbf{k}}=0$) for some nesting wavevector $\mathbf{Q}$, this is usually referred to as "perfect" particle-hole nesting. 

As noted in the previous section, in our model this occurs when $\sigma=\sigma^\prime$ and $\mathbf{Q}=(\pi,\pi)$ for the $f_{x^2-y^2}$ form factor and in the case for the $f_{xy}$ $\sigma=-\sigma^\prime$ again with the a nesting wavevector $\mathbf{Q}=(\pi,\pi)$. Assuming that this condition is satisfied for an energy window $[-\varepsilon,\epsilon]$ around the Fermi surface, we can 
approximate any of the above equations as 
\begin{align}
    \chi(\mathbf{Q})=\int_{-\varepsilon}^{\epsilon}d\xi\frac{2n_{F}(\xi)-1}{2\xi}\rho(\xi)\,.
\end{align}
Since the integrand depends on temperature via the Fermi function the we can introduce a cut-off in the integral limits as 
\begin{equation}
    \chi(\mathbf{Q})\approx\int_{-\varepsilon}^{-T}d\xi\frac{2n_{F}(\xi)-1}{2\xi}\rho(\xi)+\int_{T}^{\varepsilon}d\xi\frac{2n_{F}(\xi)-1}{2\xi}\rho(\xi)\,,
\end{equation}
where $\rho(\xi)$ is the density of states. The above equation  formally holds only in the limit $T\rightarrow0$, where the Fermi function reduces to $n_{F}(\xi)=1-\Theta(\xi)$, with $\Theta(\xi)$ being the Heaviside function. Assuming the density of states can be approximated by its value at the Fermi level $\rho(\xi)\approx\rho(0)$, which is not valid around a VHS, we get
\begin{equation}
    \chi(\mathbf{Q})=-\rho(0)\int_{T}^{\varepsilon}d\xi\frac{1}{\xi}=-\rho(0)\log\left(\frac{\varepsilon}{T}\right)\,.
\end{equation}
Therefore, this integral is guaranteed to diverge as a function of temperature, an by proxy the RPA susceptibility if the respective channel is attractive. In the case where the chemical potential lies in the vicinity of a vH singularity which imparts a logarithmic divergence in the density of states, 
\begin{equation}
     \chi(\mathbf{Q})\propto -\log^2\left(\frac{\varepsilon}{T}\right)\,,
\end{equation}
which is what we refer to as "double logarithmic scaling" in the main text. Notably, this type of scaling leads to an increase in critical temperature. A similar calculation for the  singlet particle-particle susceptibility: 
\begin{equation}
    \chi^{pp}(\mathbf{q})=\frac{1}{2}\sum_{\sigma}\int_{\mathbf{k}\in\mathrm{BZ}}\frac{1-n_{F}(\xi_{\mathbf{k}+\mathbf{q}\sigma})-n_{F}(\xi_{-\mathbf{k}-\sigma})}{\xi_{\mathbf{k}+\mathbf{q}\sigma}+\xi_{-\mathbf{k}-\sigma}} \label{eqn:pp_sus}
\end{equation}
If the single particle dispersion preserves TRS, i.e 
\begin{equation}
    \xi_{\mathbf{k}\uparrow}=\xi_{-\mathbf{k}\downarrow}\,,
    \label{eqn:pp_nesting}
\end{equation}
 Eq.(\ref{eqn:pp_sus}) for $\mathbf{Q}=0$ leads to the same
 logarithmic scaling as in the static spin susceptibility
\begin{equation}
    \chi^{pp}(0)=\rho(0)\log\left(\frac{\varepsilon}{T}\right)\,.
\end{equation}
This is commonly referred to as the "Cooper logarithm" and is a general feature of the FL, provided Eq.(\ref{eqn:pp_nesting}) holds. As described in the previous section, the altermagnetic term in the dispersion breaks TRS, therefore suppressing singlet $\mathbf{Q}=0$ superconductivity. An interesting feature of the particular model, is the perfect nesting in the pairing channel in the limit of $t=0$ i.e. the large AM regime, with $\mathbf{Q}=(\pi,\pi)$. 

\begin{equation}
    \xi_{\mathbf{k}+\mathbf{Q}\uparrow}=\xi_{-\mathbf{k}\downarrow}\,,
    \label{eqn:pp_nesting_Q}
\end{equation}
which leads to a logarithmic scaling in the particle-particle susceptibility for this particular wave vector
\begin{equation}
    \chi^{pp}(\mathbf{Q})\propto\log\left(\frac{\varepsilon}{T}\right)\,,
\end{equation}
which is an important condition for the PDW identified in our calculations.  

\subsection{Comparison with RPA}
We illustrate the effects of fluctuations and competition between orders by comparing the results presented in the main text, with phase diagrams within the Random Phase Approximation (RPA). The static renormalized susceptibility reads:
\begin{equation}
    \hat{\chi}(\mathbf{q})=\hat{\chi}^0(\mathbf{q})\left[1-\hat{U}\hat{\chi}^0(\mathbf{q})\right]^{-1}\,,
    \label{eqn:RPA}
\end{equation}

where $\hat{U}_{\alpha\beta}=\frac{U}{4}\eta_{\alpha\beta}$ with 
\begin{align}
    \eta&=
        \begin{bmatrix}
            1 & 0 & 0& 0 \\
            0 & -1 & 0& 0 \\
             0 & 0 & -1& 0 \\
              0& 0 & 0& 1 
       \end{bmatrix}.
\end{align}
and $\hat{\chi}^0$ is the bare static susceptibility given by 
\begin{equation}
    \chi^0_{\alpha\beta}(\mathbf{q},0)=-\frac{1}{2\beta}\sum_{i\nu}\int_{\mathbf{k}}\mathrm{tr}[\sigma^{\alpha}\,G(\mathbf{\mathbf{k}+\mathbf{q}},i\nu)\,\sigma^{\beta}\,G(\mathbf{k},i\nu)]\,,
\end{equation}
with $\sigma^0=\mathbb{1}$ and $\beta=1/T$ the inverse temperature.
We identify the instabilities of the Fermi-liquid state with a divergence in the above susceptibility, i.e. when the condition $U\chi^0(\mathbf{q})=1$ is satisfied.

We note, that on the level of RPA, the Hubbard interaction is repulsive in all but the spin channels. As such we cannot probe superconducting or charge instabilities without \textit{a priori} assuming the existence of an attractive interaction.

Due to the broken SU(2) symmetry on the level of the dispersion, the bare Green's function can be decomposed as $G(\mathbf{k},i\nu)_{\sigma\sigma^\prime}=G(\mathbf{k},i\nu)\mathbb{1}_{\sigma\sigma^\prime}+G^3(\mathbf{k},i\nu)\sigma^3_{\sigma\sigma^\prime}$ with 
\begin{align}
    G^{0}(\mathbf{k},i\nu)&=\frac{1}{2}\left(\frac{1}{i\nu-\varepsilon^0(\mathbf{k})-\lambda(\mathbf{k})}+\frac{1}{i\nu-\varepsilon^0(\mathbf{k})+\lambda(\mathbf{k})}\right) \\
     G^{3}(\mathbf{k},i\nu)&=\frac{1}{2}\left(\frac{1}{i\nu-\varepsilon^0(\mathbf{k})-\lambda(\mathbf{k})}-\frac{1}{i\nu-\varepsilon^0(\mathbf{k})-\lambda(\mathbf{k})}\right)
\end{align}

By substituting the expressions for the Green's functions, we immediately get its decomposition into charge and spin channels (charge: $\alpha=\beta=0$, spin: $\alpha=\beta\in\{1,2,3\}$) 
\begin{align}\label{eqn:bubbab}
    \chi^0_{\alpha\beta}(\mathbf{q})= -\frac{1}{\beta}\sum_{i\nu}\int_{\mathbf{k}}\Big\{ &G^{0}(\mathbf{k}+\mathbf{q},i\nu)G^0(\mathbf{k},i\nu)\delta_{\alpha\beta} +G^3(\mathbf{k}+\mathbf{q},i\nu)G^3(\mathbf{k},i\nu)\eta_{\alpha\beta} \nonumber   \\
   &+  [G^{0}(\mathbf{k}+\mathbf{q},i\nu)G^3(\mathbf{k},i\nu)-G^0(\mathbf{k})G^3(\mathbf{k}+\mathbf{q},i\nu)]\Phi_{\alpha\beta} \Big\}\,,
\end{align}
where $\alpha,\beta\in\{0,1,2,3\}$ refer to the generators of U(2)  and 
\begin{align}
    \Phi &= \begin{bmatrix}
            0 & 0 & 0& 1 \\
             0& 0 & i& 0 \\
             0 & -i & 0& 0 \\
              1 & 0 & 0& 0 
       \end{bmatrix}
\end{align}

\begin{figure}
    \centering
    \includegraphics[width=0.49\linewidth]{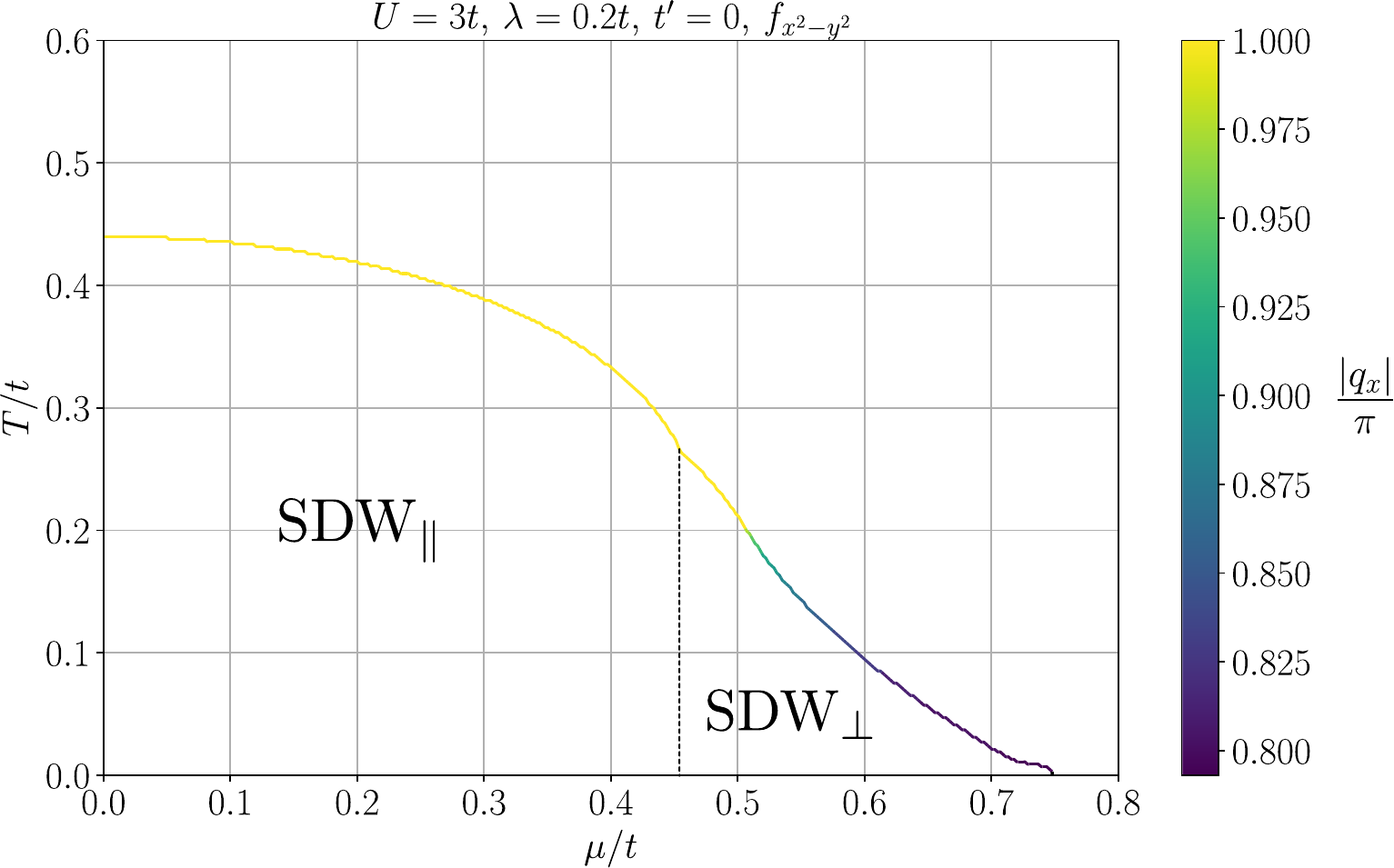}%
    \hfill
    \includegraphics[width=0.49\linewidth]{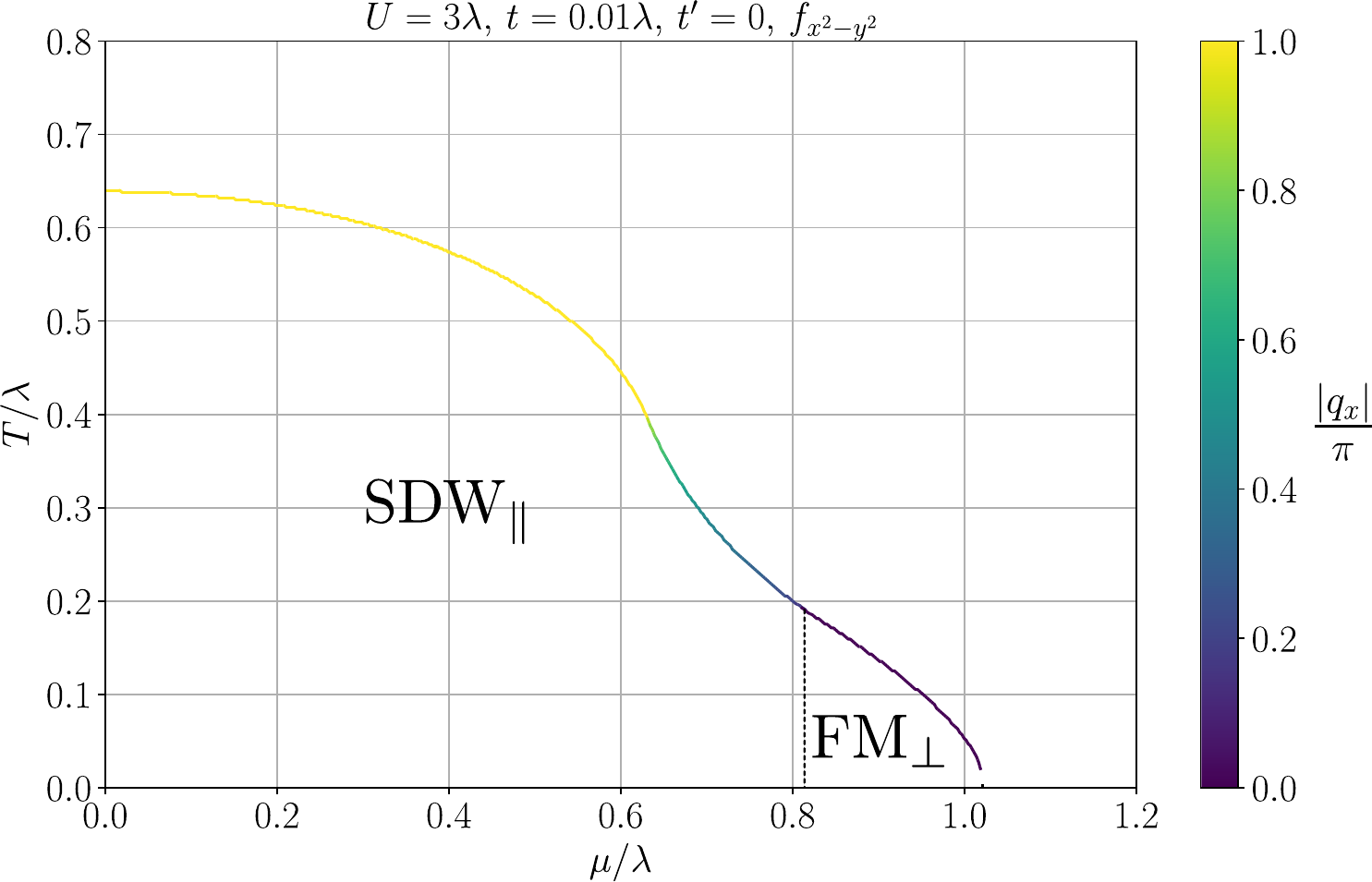}
    \includegraphics[width=0.49\linewidth]{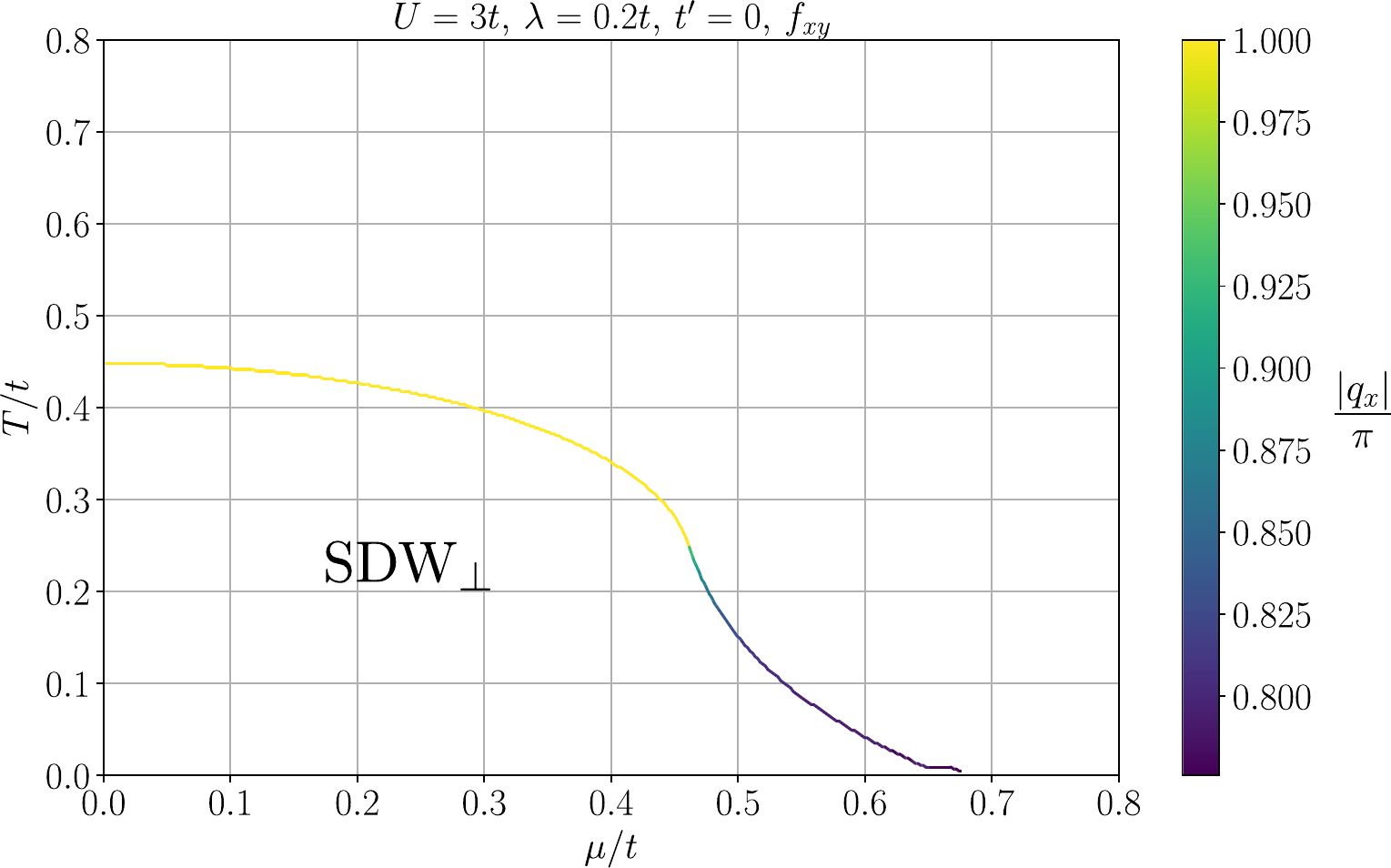}%
    \hfill\includegraphics[width=0.49\linewidth]{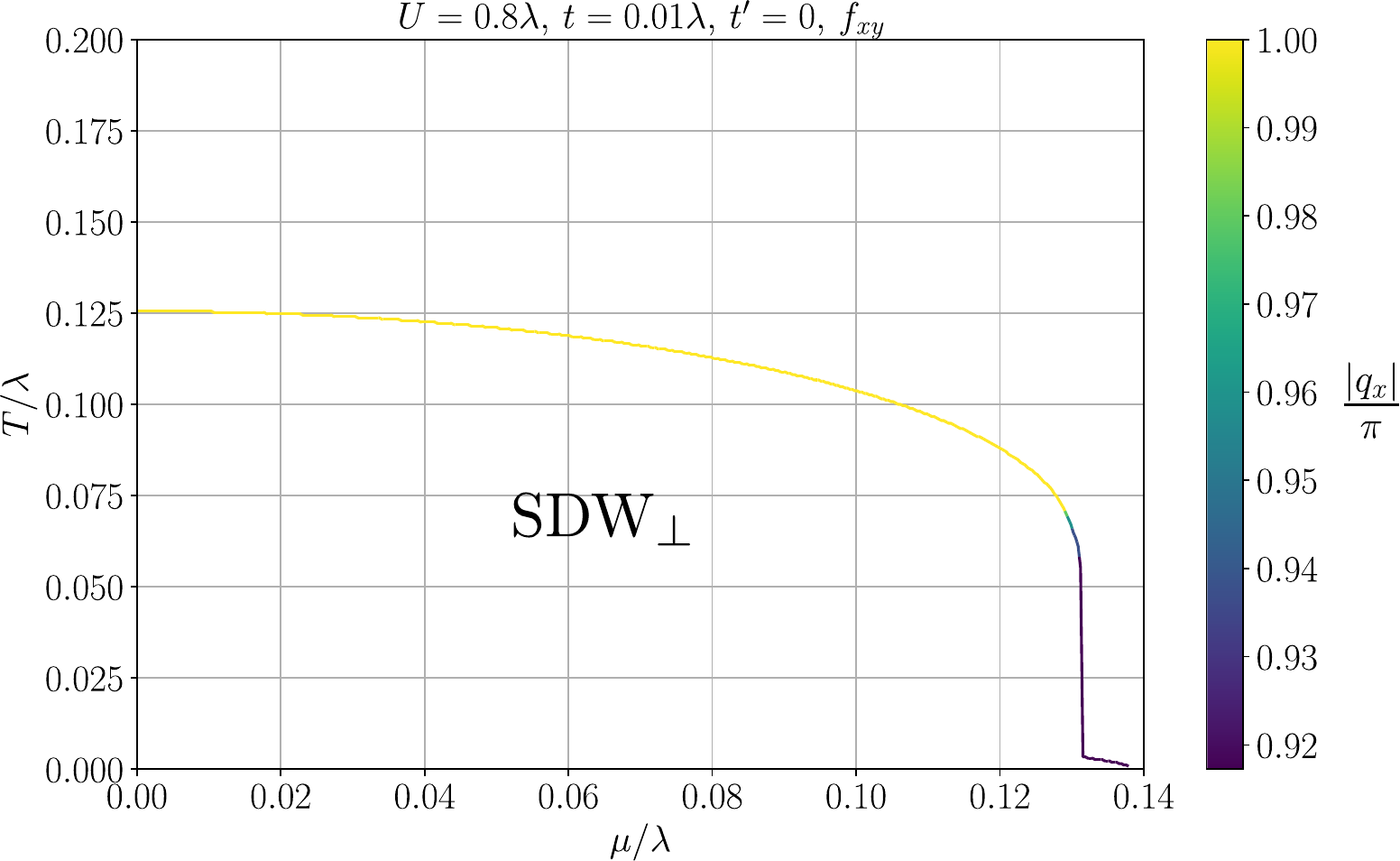}
    \caption{RPA phase diagrams as function of chemical potential $\mu$ and temperature $T$. Top: $f_{x^2-y^2}$ Bottom: $f_{x^2-y^2}$. The color scale encodes the commensurability of the magnetic state. The dashed line indicates the chemical potential defining the phase boundary between the magnetic phases. The results are qualitatively similar to the FRG. Since RPA does not include inter-channel feedback, the critical scales are overestimated.}
    \label{fig:SM_RPA}
\end{figure}

The term multiplying the $\Phi$ matrix can be shown to vanish, by shifting the integration variable to $\mathbf{k}\rightarrow-\mathbf{k}-\mathbf{q}$ due to $G^0, G^3$ being even functions of $\mathbf{k}$.
For the set of parameters studied we check where Eq.~(\ref{eqn:RPA}) exhibits a pole, which we identify with the onset of a collective excitation with wavevector $\mathbf{q}$. We provide several phase diagrams for different parameters in Fig.~(\ref{fig:SM_RPA}) for comparison with the FRG results provided in the main text. As can be seen in Fig.~\ref{fig:SM_RPA} the critical temperature for the magnetic states is overestimated as the inter-channel feedback is absent. We also note the order can be incommensurate as can be seen by the color code.

\section{PDW pairing mechanism}
\label{supp:PDW_glue}
In this section we elucidate on the mechanism enabling the attractive interaction in the PDW identified with an extended s-wave form factor in the triplet (symmetric) spin configuration. Due to both the triplet character of this state and the fact it carries a finite momentum 
$\mathbf{q}=\mathbf{M}$, the understanding of the pairing glue cannot directly be inferred from the "usual" case of spin-fluctuation mediated d-wave superconductivity. We will thus analyze the structure of its flow equations to gain analytical insight into the relevant fluctuations that mediate the attraction.

We reiterate that the pairing instability occurs for the symmetric triplet combination with $t^3=i\tau^2\tau^3$, where $\tau_i$ are spin Pauli matrices, and with form factor $f^1_\mathbf{k}=\cos{k_x}+\cos{k_y}$. As such within our notation is labeled $P^{33}_{11}(\mathbf{q})$. 
Its flow equation reads:

\begin{equation} 
\begin{aligned}
    \Dot{P}^{33}_{11}(\mathbf{q})=
    &-\frac{1}{2}\sum_{\alpha',\beta'}\sum_{m,m'}P[V]^{3\beta'}_{1 m'}(\mathbf{q})\Dot{\Pi}^{\beta\alpha'}_{P,mm'}(\mathbf{q})P[V]^{\alpha'3}_{m'1}(\mathbf{q})\,,
\end{aligned}
\end{equation}
where the superscript indexes the t- matrix, the subscript the form-factor components and $P[V]^{3\beta'}_{1 m'}(\mathbf{q})$ is the projection of the vertex on the particle-particle channel. This is the crucial point of the FRG, since the vertex contains all contributions from quantum fluctuations in an unbiased way. This is especially important for electronically mediated superconductivity, arising from particle-hole fluctuations, akin to the Kohn-Luttinger mechanism. . 

Due to $P^{33}_{11}(\mathbf{q})$ being the divergent component in the particle-particle channel, in our simplified analytic approch to the flow equations we 
treat the others as negligible by setting them to zero. This amounts to only retaining the respective diagonal components for the projections and $\Dot{\Pi}$. Similarly, in the present scenario, in the vicinity of the PDW instability in the phase diagram, only the transverse ferromagnet (FM$_\perp$) and the longitudinal spin-density wave (SDW$_\|$) are identified as instabilities. Thus, we expect  these to be the dominant contributions in the particle-hole channel and
would like to investigate if their fluctuations 
are responsible for the PDW pairing glue. As such we only keep the $D^{11}_{00}(\mathbf{q}), D^{22}_{00}(\mathbf{q})$ and $D^{33}_{00}(\mathbf{q})$ contributions in the particle-hole channel and  approximate the total vertex as 
\begin{equation}
\begin{aligned}
    V^{\sigma_1\sigma_2\sigma_3\sigma_4}(\mathbf{k}_1,\mathbf{k}_2,\mathbf{k}_3;T)&\approx U(\delta_{\sigma_1\sigma_3}\delta_{\sigma_2\sigma_4}-\delta_{\sigma_1\sigma_4}\delta_{\sigma_2\sigma_3})
     +\frac{1}{2}(P^{33}_{11}(\mathbf{k}_1+\mathbf{k}_2)t^3_{\sigma_1\sigma_3}(t^3)^\dagger_{\sigma_2\sigma_4}f^1_{\mathbf{k}_3}f^{1}_{\mathbf{k}_4}\\&
     +D^{11}_{00}(\mathbf{k}_3-\mathbf{k}_1)f^{0}_{\mathbf{k}_1}f^0_{\mathbf{k}_4}\tau^1_{\sigma_1\sigma_3}\tau^1_{\sigma_2\sigma_4}
     +D^{22}_{00}(\mathbf{k}_3-\mathbf{k}_1)f^{0}_{\mathbf{k}_1}f^0_{\mathbf{k}_4}\tau^2_{\sigma_1\sigma_3}\tau^2_{\sigma_2\sigma_4}\\
     &+D^{33}_{00}(\mathbf{k}_3-\mathbf{k}_1)f^{0}_{\mathbf{k}_1}f^0_{\mathbf{k}_4}\tau^3_{\sigma_1\sigma_3}\tau^3_{\sigma_2\sigma_4}
     -D^{11}_{00}(\mathbf{k}_2-\mathbf{k}_3)f^{0}_{\mathbf{k}_1}f^0_{\mathbf{k}_3}\tau^1_{\sigma_1\sigma_4}\tau^1_{\sigma_2\sigma_3}\\
     &-D^{22}_{00}(\mathbf{k}_2-\mathbf{k}_3)f^{0}_{\mathbf{k}_1}f^0_{\mathbf{k}_3}\tau^2_{\sigma_1\sigma_4}\tau^2_{\sigma_2\sigma_3}\\
     &-D^{33}_{00}(\mathbf{k}_2-\mathbf{k}_3)f^{0}_{\mathbf{k}_1}f^0_{\mathbf{k}_3}\tau^3_{\sigma_1\sigma_4}\tau^3_{\sigma_2\sigma_3})\,,
\end{aligned}
\end{equation}
where $f^0=1$.
We perform the projections 

\begin{equation}
    P^{33}_{11}[V](\mathbf{q})=\frac{1}{2}\int_{\mathbf{k}\mathbf{k}'}\sum_{\substack{\sigma_1,\sigma_2\\\sigma_3,\sigma_4}} V^{\sigma_1\sigma_2\sigma_3\sigma_4}(\mathbf{k},\mathbf{q}-\mathbf{k},\mathbf{k}')(t^3)^\dagger_{\sigma_1\sigma_2}t^3_{\sigma_3\sigma_4}f^1_{\mathbf{k}}f^1_{\mathbf{k}'}
\end{equation}

For the channel to be attractive within our convention, $\Dot{P}$ has to be negative, meaning that the overall the contribution from the projections should yield a positive sign, which depends on the spin contractions and form factor integrations in order for the channel to grow along the flow. We calculate the contributions explicitly for the fluctuations in the particle-hole channel
\begin{equation}
    \begin{aligned}
       D_{d} & =\frac{1}{4}\int_{\mathbf{k}\mathbf{k}'}\sum_{\substack{\sigma_1,\sigma_2\\\sigma_3,\sigma_4}} D^{\alpha\alpha}_{00}(\mathbf{k}-\mathbf{k}')\tau^{\alpha}_{\sigma_1\sigma_3}\tau^{\alpha}_{\sigma_1\sigma_4}f^0_{\mathbf{k}}f^0_{\mathbf{q}-\mathbf{k}'}(t^3)^\dagger_{\sigma_1\sigma_2}t^3_{\sigma_3\sigma_4}f^1_{\mathbf{k}}f^1_{\mathbf{k}'}\\
        D_{ex} & =-\frac{1}{4}\int_{\mathbf{k}\mathbf{k}'}\sum_{\substack{\sigma_1,\sigma_2\\\sigma_3,\sigma_4}} D^{\alpha\alpha}_{00}(\mathbf{q}-\mathbf{k}-\mathbf{k}')\tau^{\alpha}_{\sigma_1\sigma_3}\tau^{\alpha}_{\sigma_1\sigma_4}f^0_{\mathbf{k}}f^0_{\mathbf{k}'}(t^3)^\dagger_{\sigma_1\sigma_2}t^3_{\sigma_3\sigma_4}f^1_{\mathbf{k}}f^1_{\mathbf{k}'}\,.
    \end{aligned}
\end{equation}
Note that $\mathbf{q}$ specifies that momentum structure of the projected channel. In this case the PDW exhibits peaks at $\mathbf{q}=\mathbf{M}=(\pi,\pi)$. The subscript $d/ex$, stands for direct (exchange) particle-hole channel contributions in the full vertex, which takes care of the antisymmetrization. The superscript indexes the three different SDW channels $\alpha=\{1,2,3\}$. 
Since all the spin indices are summed over, we end up with a trace over Pauli matrices
\begin{equation}
    \sum_{\substack{\sigma_1,\sigma_2\\\sigma_3,\sigma_4}}\tau^{1}_{\sigma_1\sigma_3}\tau^{1}_{\sigma_1\sigma_4}(t^3)^\dagger_{\sigma_1\sigma_2}t^3_{\sigma_3\sigma_4}=\mathrm{Tr}[\tau^1\tau^1\tau^1\tau^1]=2.
\end{equation}
Similarly, for $D^{22}_{00}$, $\mathrm{Tr}[\tau^1\tau^2\tau^1\tau^2]=2$ and for $D^{33}_{00}$, $\mathrm{Tr}[\tau^1\tau^3\tau^1\tau^3]=-2$. Note that the longitudinal SDW enters with an opposite sign compared to the transversal FM contributions. To calculate the integrals analytically, we approximate the channels as $\delta$-functions with peaks at the wave vectors where they diverge. This is valid since we expect  these fluctuations to drive the attraction. We thus have
\begin{align}
    D^{11}_{00}(\mathbf{x})=D^{22}_{00}(\mathbf{x})=D^{\perp}_{00}(\mathbf{x})&=\delta(\mathbf{x})\\
    D^{33}_{00}(\mathbf{x})=D^{\|}_{00}(\mathbf{x})&=\delta(\mathbf{x}-\mathbf{M})\,.
\end{align}
Since the longitudinal channel exhibits peaks at $(\pi,\pi)$, performing the integral over  $\mathbf{k}'$ sets $\mathbf{k}'=\mathbf{k}+\mathbf{M}$, which sets $f^1_{\mathbf{k}'}=-f^1_{\mathbf{k}}$, changing the sign of the longitudinal contribution. This is analogous to the well-known mechanism how SDWs can mediate an attraction by connecting regions of opposite sign in the gap function via the SDW wave vector. The integral $\int_\mathbf{k}(f^1_{\mathbf{k}})^2=2$ over the BZ yields a positive contribution. 
In the ferromagnetic case, $D^{11}_{00}$ and  $D^{22}_{00}$ peak at zero, setting $\mathbf{k}'=\mathbf{k}$. 
Therefore the sign of these terms is positive as well, i.e., FM fluctuations mediate pairing in a triplet channel as expected. The opposite happens for the $D_{ex}$ contributions since they come with a $\mathbf{q}-\mathbf{k}-\mathbf{k}'$ in the integrand, that is, $\mathbf{k}'=\mathbf{k}+\mathbf{M}$ in the ferromagnetic channel and $\mathbf{k}'=\mathbf{k}$ in the longitudinal one. As such, all terms yield an overall positive contribution and both channels contribute in generating an attractive interaction for this pairing instability.  

To understand why the extended $s$-wave form factor corresponds to the leading pairing instability, we compare the critical temperatures of other possibilities. For the form factors considered in our FRG scheme (see Methods) a finite $\mathbf{q}=\mathbf{M}$ pairing state is only allowed for $f^{1}_{\mathbf{k}}=\cos k_x+\cos k_y$ and  $f^{2}_{\mathbf{k}}=\cos k_x-\cos k_y$. The fact that the dominant instability is the one corresponding to $f^{1}_{\mathbf{k}}$ can be understood by its nodal structure. For $t=0$, the two spin split FS are related by $\xi_{\mathbf{k}+\mathbf{M}\sigma}=\xi_{-\mathbf{k}\Bar{\sigma}}$. In this case the dispersion relation fully inherits the nodal structure of the AM form-factor. Looking at the linearized gap equation  
\begin{align}
  \sum_{\mathbf{k'}}V_{\mathbf{k},\mathbf{k'}}\frac{\tanh{(\frac{\xi_{\mathbf{k'}\sigma}}{2T})}}{2\xi_{\mathbf{k'}\sigma}}\Delta_\mathbf{k'}= \Delta_\mathbf{k}\,,
  \label{eqn:gapeq}
\end{align}
where $V_{\mathbf{k},\mathbf{k'}}$ labels the effective attractive interaction for the respective channel. If $\Delta_\mathbf{k}=\Delta_0f^{2}_{\mathbf{k}}$, it perfectly cancels the denominator of Eq.~(\ref{eqn:gapeq}), thus suppressing the logarthmic divergence of the kernel at $T=0$. In contrast, in the case $\Delta_\mathbf{k}=\Delta_0f^{1}_{\mathbf{k}}$, this does not occur, thus promoting this particular nodal structure. For a finite $t$, $\xi_{\mathbf{k}+\mathbf{M}\sigma}=\xi_{-\mathbf{k}\Bar{\sigma}}$ no longer holds, However we expect for small enough values of $t$ for this effect to still persist. Indeed as can be seen in Fig.~4d of the main text, the PDW still occurs until $t\sim0.04$. 

\section{Additional FRG phase diagrams}
\begin{figure*}[ht!]
    \centering
    \includegraphics[width=\linewidth]{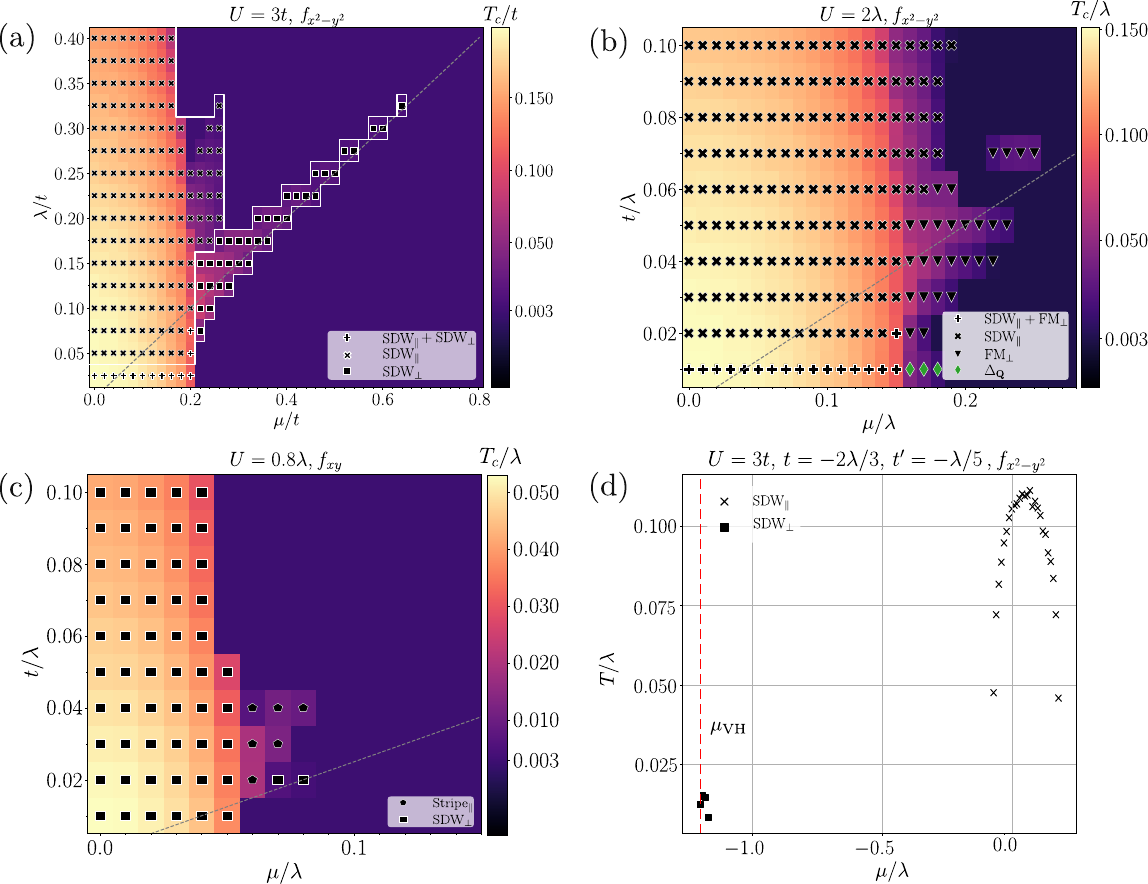}
    \label{fig:rest_frg}
    \caption{Top: FRG phase diagrams as function of chemical potential $\mu$ and the ratio of hopping and AM splitting $t/\lambda$ ($\lambda/t$) for the small (large) AM regime in the $f_{x^2-y^2}$ case for different values of the interaction $U$. The color scale encodes the critical temperature $T_c$ and the dashed line indicates VH filling. 
    Bottom: analogous FRG phase diagram for the large AM regime in the  $f_{xy}$ case. 
    }
\end{figure*}

In this section we provide the FRG diagrams not included in the main text. 
For the $f_{x^2-y^2}$ form factor, in the small AM regime, the splitting $\lambda$ affects the location of the vH singularity as a function of $\mu$. This splits the phase diagrams into two regions, one dominated by SDW$_\|$ around $\mu=0$ due to perfect same spin nesting and one by SDW$_\perp$ enabled by the spin-split vH singularity at $\mu=2\lambda$ (Fig.~\ref{fig:rest_frg}). For the the $f_{xy}$ form factor in the large AM regime, the instability landscape is dominated by SDW$_\perp$, driven by opposite spin nesting for $\mu\lesssim 0.05\lambda$ (Fig.~\ref{fig:rest_frg}). A small region ($0.06\lambda\lesssim\mu\lesssim 0.08\lambda$ and  $0.02\lambda\lesssim t\lesssim 0.04\lambda$ is dominated by a longitudinal stripe order is found in the vicinity of $t\sim0$, where perfect same spin is present for $\mathbf{Q}=(\pi,0)$. 

\begin{figure}[ht!]
    \centering
    \includegraphics[width=0.5\linewidth]{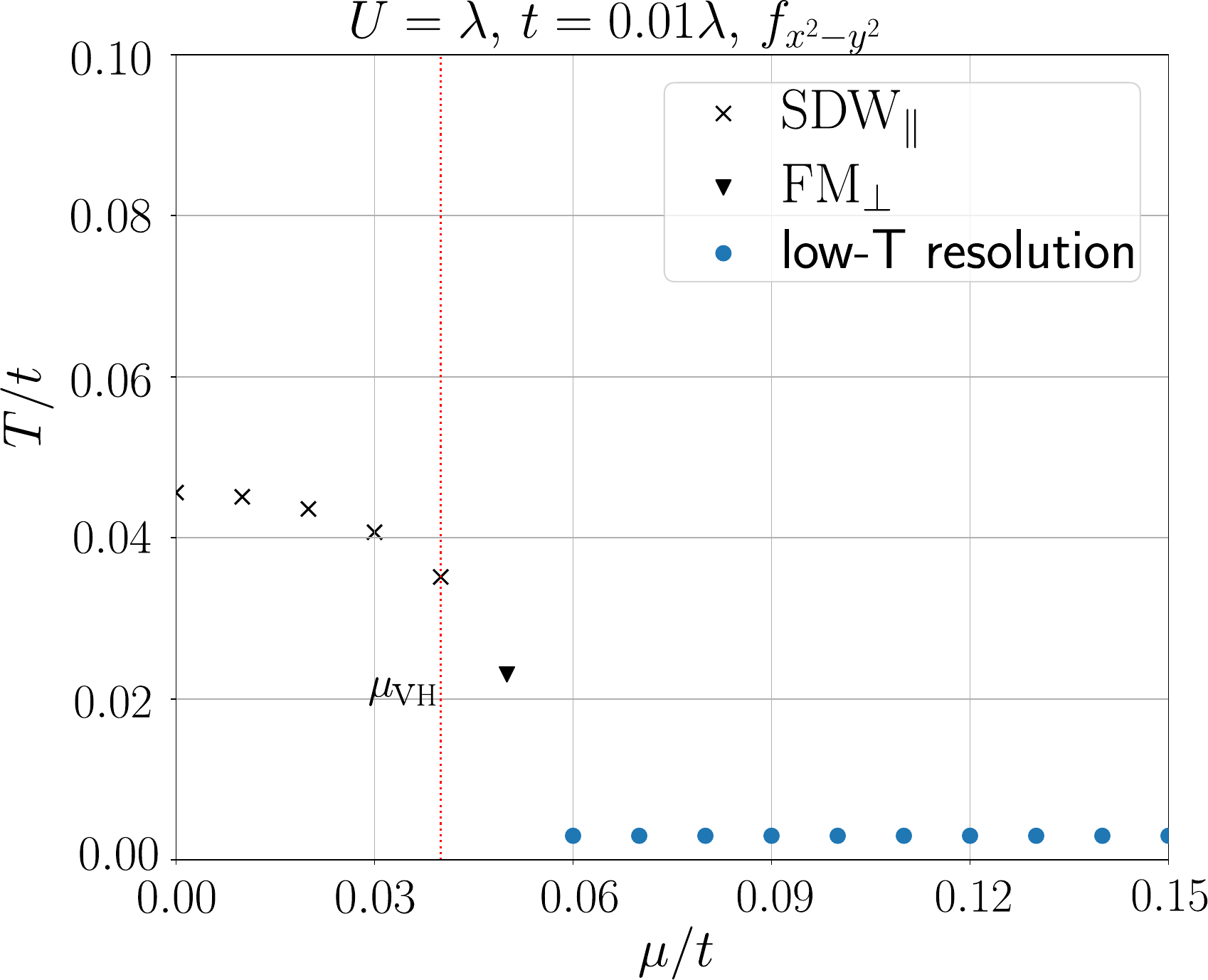}%
    \includegraphics[width=0.5\linewidth]{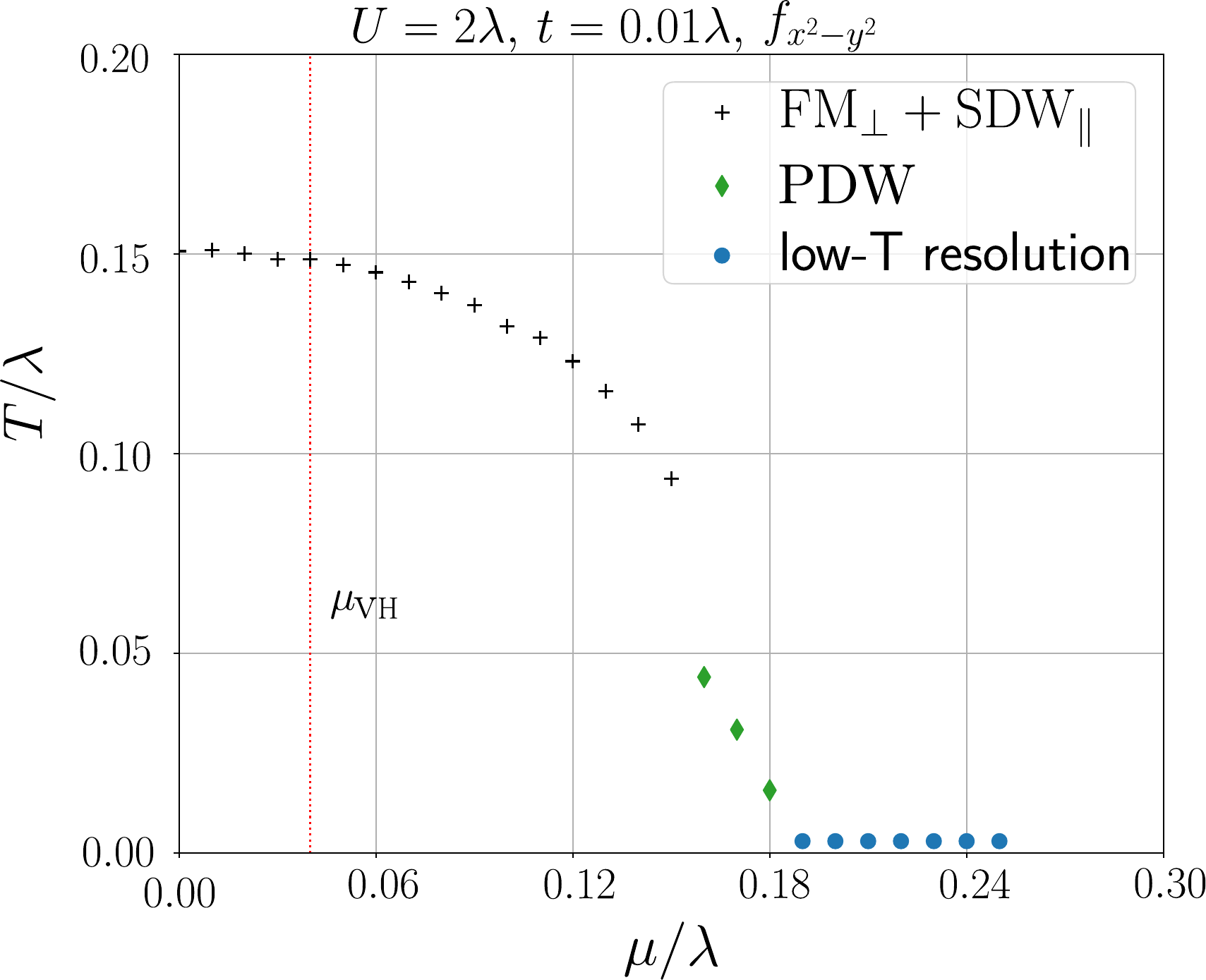}
    \caption{FRG 
    critical temperature $T_c$ as function of chemical potential $\mu$ of a large $f_{x^2-y^2}$ AM splitting for different values of the bare Hubbard $U$.}
    \label{fig:SM_U_Dep}
\end{figure}

As noted in the main text, the FRG phase diagrams have a non-universal dependence on the value of the bare Hubbard $U$. This occurs due to the inter-channel feedback inherent to the FRG flow equations. More specifically, for the $f_{x^2-y^2}$ form factor in the large AM regime, this leads to a boost of the transverse channel, where a ferromagnetic instability occurs (Fig.~\ref{fig:SM_U_Dep}). 
For smaller values of $U$, the critical scales are significantly reduced and the instabilities occur due to the logarithmic divergences due to (perfect) nesting and the presence of VH singularities.